\documentclass[12pt]{article}
\usepackage{amssymb,amsmath}
\usepackage{epsfig,psfrag}
\usepackage{curves}
\usepackage{amsmath}
\usepackage{graphicx}
\usepackage{color}
\definecolor{darkblue}{cmyk}{0.9,0.9,0,0}
\usepackage[colorlinks=true,linkcolor=darkblue,citecolor=darkblue,urlcolor=darkblue]{hyperref}
\usepackage{epsfig}

\newcommand{\be}{\begin{equation}}
\newcommand{\ee}{\end{equation}}
\newcommand\beqa{\begin{eqnarray}}
\newcommand\eeqa{\end{eqnarray}}

\def\XXint#1#2#3{{\setbox0=\hbox{$#1{#2#3}{\int}$}
\vcenter{\hbox{$#2#3$}}\kern-.5\wd0}}

\newcommand{\nn}{\nonumber}

\newcommand{\neqa}{\nonumber\end{eqnarray}}
\newcommand{\la}[1]{\label{#1}}

\def\tr{{\rm tr~}}

\newcommand{\Tr}{{\rm Tr}}

\renewcommand{\d}{\partial}

\newcommand{\<}{{\langle}}
\renewcommand{\>}{{\rangle}}

\renewcommand{\sp}{p\hspace{-.40em}/}

\def\su2{{SU(2)}}

\def\[{\left[}
\def\]{\right]}

\def\e{\epsilon}

\def\({\left(}
\def\){\right)}
\def\[{\left[}
\def\]{\right]}

\def\<{\langle}
\def\>{\rangle}

\def\i2{\frac{i}{2}}

\def\spi{\relax{\rm \pi\kern-0.5em /}}
\def\sA{\relax{\rm A\kern-0.5em /}}
\def\sp{\relax{\rm p\kern-0.5em /}}
\def\sd{\relax{\rm \d\kern-0.5em /}}
\def\sk{\relax{\rm k\kern-0.5em /}}
\def\sn{\relax{\rm n\kern-0.5em /}}
\def\sl{\relax{\rm l\kern-0.5em /}}
\def\sP{\relax{\rm P\kern-0.7em /}}
\def\sBethe{\relax{\rm \Bethe\kern-0.5em /}}
\def\cN{{\cal N}}

\usepackage{makeidx}
\makeindex

\usepackage[english]{babel}

\catcode`\@=11
\def \Tr {\mathop{\rm Tr}\nolimits}
\def \tr {\mathop{\rm tr}\nolimits}

\def \e  {\mathop{\rm e}\nolimits}
\newcommand\lr[1]{{\left({#1}\right)}}

\newcommand\re[1]{(\ref{#1})}

\newcommand{\pa}{\partial}

\renewcommand{\d}{\delta}

\newcommand{\ep}{\epsilon}

\newcommand{\p}[1]{(\ref{#1})}
\newcommand{\bt}[1]{{\bar t}}
\newcommand{\ts}{\textstyle}

\newcommand{\half}{{\ts \frac{1}{2}}}
\newcommand \vev [1] {\langle{#1}\rangle}

\newcommand {\cO}{{{\cal O}}}

\newcommand{\beq}{\begin{equation}}
\newcommand{\eeq}{\end{equation}}
\newcommand{\bea}{\begin{eqnarray}}
\newcommand{\eea}{\end{eqnarray}}
\newcommand{\ena}{\end{eqnarray}}
\newcommand{\bear}{\begin{eqnarray}}
\newcommand{\ear}{\end{eqnarray}\noindent}

\def\nref#1{(\ref{#1})}

\newcommand{\ft}[2]{{\textstyle\frac{#1}{#2}}}

\def\numberbysection{\@addtoreset{equation}{section}
                     \def\theequation{\thesection.\arabic{equation}}}
\numberbysection

\begin{document}


\thispagestyle{empty}

\null\vskip-12pt \hfill
\begin{minipage}[t]{35mm}
 LU-ITP 2010/002 \\
IPhT--T10/91 \\
 LAPTH--024/10
\end{minipage}

\vskip3.2truecm
\begin{center}
\vskip 0.2truecm {\Large\bf
From correlation functions to Wilson loops}

\vskip 1truecm

{\bf   Luis F. Alday$^{*}$, Burkhard Eden$^{**}$, Gregory P. Korchemsky$^{\dagger}$,\\ Juan Maldacena$^{*}$, Emery Sokatchev$^{\ddagger}$ \\
}

\vskip 0.4truecm
$^{*}$ {\it School of Natural Sciences, Institute for
Advanced Study,\\
 Princeton, NJ 08540, USA\\
 \vskip .2truecm
$^{**}$ Institut f\"ur theoretische Physik, Universit\"at Leipzig \\
Postfach 100920, D-04009 Leipzig, Germany \\
 \vskip .2truecm
$^{\dagger}$ Institut de Physique Th\'eorique\,\footnote{Unit\'e de Recherche Associ\'ee au CNRS URA 2306},
CEA Saclay, \\
91191 Gif-sur-Yvette Cedex, France\\
\vskip .2truecm $^{\ddagger}$ LAPTH\,\footnote[2]{Laboratoire d'Annecy-le-Vieux de Physique Th\'{e}orique, UMR 5108},   Universit\'{e} de Savoie, CNRS, \\
B.P. 110,  F-74941 Annecy-le-Vieux, France
                       } \\
\end{center}

\vskip 1truecm 
\centerline{\bf Abstract} 
\medskip
\noindent

We start with an  $n-$point correlation function in a conformal
gauge theory. We show that a special limit produces a polygonal
Wilson loop with $n$ sides. The limit takes  the $n$ points towards
the vertices of a null polygonal Wilson loop such that successive
distances $x^2_{i,i+1} \to 0$. This produces a fast moving particle
that generates a ``frame'' for the Wilson loop. We explain in detail
how the limit is approached, including some subtle effects from the
propagation of a fast moving particle in the full interacting
theory. We perform perturbative checks by doing explicit
computations in ${\cal N}=4$ super-Yang-Mills theory.

\newpage

\thispagestyle{empty}

{\small \tableofcontents}

\newpage
\setcounter{page}{1}\setcounter{footnote}{0}



\section{Introduction}

The natural observables in a conformal field theory are correlation functions of gauge invariant local operators,
\begin{equation}\label{nO}
   G_n= \vev{\cO(x_1)\cO(x_2)\dots \cO(x_n)}\,.
\end{equation}
These correlation functions have well-controlled singularities when two points approach each other, $x_i \to x_j$,
along a space-like direction. This leads to the well-understood Euclidean operator product expansion (OPE)
in terms of local operators.
In many applications it is interesting to consider a limit in which the proper distance between two points
becomes zero, $x_{ij}^2 \to 0$, but the points  remain separated along a null direction. This leads to the
well studied
 light-cone OPE
 which is important for many high energy processes
in QCD, such as deep inelastic scattering, etc.

\begin{figure}[h!]
\psfrag{(a)}[cc][cc]{(a)}\psfrag{(b)}[cc][cc]{(b)}\psfrag{(c)}[cc][cc]{(c)}
\centerline{ \includegraphics[height=35mm]{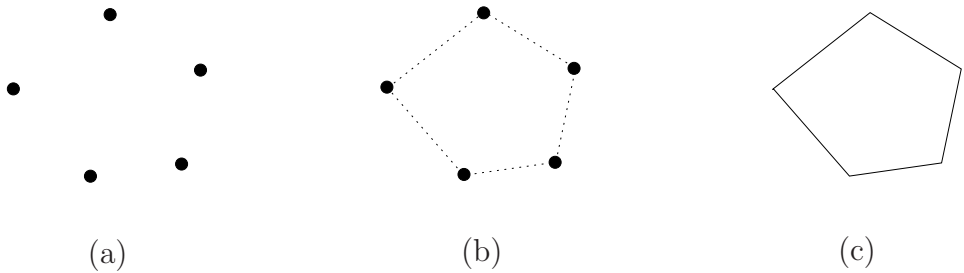} }
 \caption{ \small  (a) Diagrammatic representation of the correlation function; the black dots denote the points $x_i$. (b) The distances $x^2_{i,i+1}$ go to zero. (c) We are left
 with a Wilson loop on a polygonal contour with null sides.
 \label{corrwislon}}
 \end{figure}

In this paper we consider a situation where the points $x_i$ ($i=1,\ldots,n$) become light-like separated from each other in a
sequential fashion,
$x_{i,i+1}^2 \to  0$ (with the cyclic identification $x_{i+n} \equiv x_i$), so that
many distances are becoming light-like at the same time.
We find that in this limit the leading asymptotic behavior of the correlator $G_n$ is given by 
{ the product of a tree-level correlator and} the expectation value of a polygonal Wilson loop operator, defined on a piecewise
  null polygon $C_n$ with cusps at the points $x_i$,
  in the adjoint representation of the gauge group $SU(N_c)$:
\begin{align}\label{Wl0}
 W^{\rm adj}[C_n]=\frac1{N_c^2-1}\vev{0|\Tr_{\rm adj} {\rm P}\exp \lr{ig\oint_{C_n} dx\cdot  A(x) }|0}\,.
\end{align}
Taking the limit of the correlator gives
\begin{equation}\label{cowll0}
    \lim_{x^2_{i,i+1} \to 0} G_n / G_n^{\rm tree}  \propto  W^{\rm adj}[C_n] \,,
\end{equation}
where the proportionality factor depends on the details of the limit and we will discuss it further below.
Here we quoted the result for a theory whose fields are all in the adjoint representation of the gauge group. In the planar
limit this can be approximated by a product of a Wilson loop in the fundamental and another one in the anti-fundamental { representation of $SU(N_c)$}.

In this way we find a concrete relation between
correlation functions of local operators and certain Wilson loop operators.
We motivate the connection and provide
 explicit evidence for the statements we make, in the case of ${\cal N}=4 $ super-Yang-Mills theory (SYM) in the  planar limit. However, the statements regarding the emergence of the Wilson loop should be valid for a general conformal field theory in any space-time dimension.

The basic reason why we obtain the Wilson loop is very simple. When two points are null separated, $x_{i,i+1}^2\to 0$,
there is a singularity in the correlator that arises from a { very fast } particle propagating  between
these points. In a free theory, this is the ordinary $1/x_{i,i+1}^2$ singularity of the propagator.
In the interacting theory, the fast particle going around the various vertices of the polygon
is charged under the gauge group, so it interacts with the gauge field. When $x_{i,i+1}^2 \to 0$, the propagator of the interacting particle  can be approximated by
a light-like Wilson line connecting the  points $x_i$ and $x_{i+1}$. The full approach to the limit is a bit  subtle, since one has to carefully regularize the two singular objects  in \p{cowll0}. Our goal is to explain the limit in detail and to provide several explicit checks of the correctness of the
arguments.

The light-cone limit of the correlator can be performed in different ways. The first is to start with $G_n$ in four dimensions, with space-like separated points. There the correlator is well defined and manifestly conformally covariant. Then one approaches the limit
where $x_{i,i+1}^2 \to 0$ and the points define a polygonal loop with null sides.
In taking this limit one gets leading  divergent terms of the type
\begin{equation}
G_n/G_n^{\rm tree} \sim  \exp\left( - { \Gamma^{\rm adj}_{\rm cusp}  \over 4} \sum_{i=1}^n \log x_{i,i+1}^2 \log x_{i-1,i}^2 \right)\,,
\end{equation}
where $\Gamma^{\rm adj}_{\rm cusp} $ is the cusp anomalous dimension   in the adjoint representation of the
gauge group. It is   known to determine the leading UV singularity of   light-like   Wilson loops
\cite{P80,KR87,KK92}\footnote{Our definition of the cusp anomalous dimension differs by a factor
of two from the one in \cite{P80,KR87,KK92} and other papers in the literature,
$\Gamma_{\rm cusp}^{\rm here} = 2 \Gamma_{\rm cusp}^{\rm there} $.},
as well as the IR singularity of amplitudes \cite{Sen:1981sd,KR86,IKR85}.
In the present context,
 the cusp anomalous dimension  also gives the
 leading short-distance singularity of correlators.
 Thus  the distances $x_{i,i+1}^2$ serve as a UV regulator for
the Wilson loop. In fact, they regulate
the { Wilson} loop in a rather subtle way, since there are further logarithmic terms whose structure depends on
more details about the particles propagating along the { polygon contour}. These details are irrelevant if we only
do a one-loop perturbative computation, but they become important when we go to two loops and higher. We
elucidate the full structure of the approach to the limit. The final formula is given in \nref{finexpan} {below}.

Another way to take the limit is to first regularize the theory {in the ultraviolet} by, e.g., formulating it in $D=4-2\ep$ dimensions (with $\ep>0$). This simplifies the approach to the limit  $x^2_{i,i+1}=0$.
 To be more precise, we take the limit so that these distances remain much smaller than the UV regulator scale, $x^2_{i,i+1} \ll 1/\mu^2$. This  is a well-defined procedure
in dimensional regularization. It requires, however, the
computation of the correlator in $D=4-2\ep$ dimensions. From a practical point of view, this is harder than the computation in $D=4$ dimensions. From a conceptual point of view, the advantage is that this approach to
the limit is much simpler because the theory becomes free in the UV regime. The particle propagators have $1/x_{i,i+1}^2$ { poles} with no further corrections. Thus, the contribution from these propagators is the same as the one present in the tree-level correlator
\begin{equation}\label{cowll}
    \lim_{x^2_{i,i+1} \to 0} { G_n \over  G^{\rm tree}_n } =    W^{\rm adj}_ \epsilon[C_n] \,,
\end{equation}
where both the left- and right-hand sides have been defined in the dimensionally regulated theory.
Of course, this would also be the behavior  in any theory that is UV free, such as  $(2+1)-$dimensional Yang-Mills theory.

The  polygonal Wilson loops  \p{Wl0} have been intensively studied during  the past
few years due to their connection with scattering amplitudes
in ${\cal N}=4$ super-Yang-Mills
\cite{am07,Drummond:2007aua,Brandhuber:2007yx,Drummond:2007cf,Alday:2007he,Drummond:2007au,Bern:2008ap,Drummond:2008aq,Drummond:2008vq,Berkovits:2008ic,Beisert:2008iq}. The present paper was also motivated by this study. Planar ${\cal N}=4$ SYM theory
is integrable and the integrability might allow us to compute either Wilson loops or correlation functions.
The connection we propose here can be used either as a way to extract a Wilson loop from a known correlator, or
as a way to constrain an unknown correlator with the knowledge of the Wilson loop expectation value.

This paper is organized as follows.
 First we discuss the dimensionally
 regularized version of the statement \nref{cowll} in section \ref{dimregsection}.
 We perform one-loop checks of this statement in section \ref{explicit}.
In section \ref{Wlfcf} we take the limit of the full four dimensional correlation function and
derive the precise approach to the limit that produces the Wilson loop. Finally, we end with some conclusions.
The paper contains several appendices addressing various  technical issues.


\section{From correlators to Wilson loops in dimensionally\\ regularized theories}
\label{dimregsection}

\subsection{A short explanation}

\begin{figure}[h!]
\psfrag{dots}[cc][cc]{$\mathbf{\dots}$}
\psfrag{1}[cc][cc]{$x_1$}
\psfrag{2}[cc][cc]{$x_2$}
\psfrag{3}[cc][cc]{$x_3$}
\psfrag{4}[cc][cc]{$x_4$}
\psfrag{n-1}[cc][cc]{$x_{n-1}$}
\psfrag{n}[cc][cc]{$x_n$}
\psfrag{(a)}[cc][cc]{(a)}
\psfrag{(b)}[cc][cc]{(b)}
%
\centerline{ \includegraphics[height=50mm]{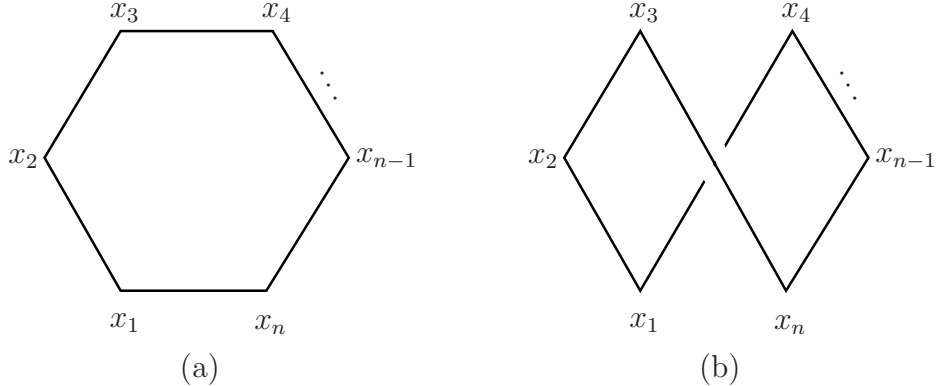} }
 \caption{\small  Feynman diagrams of different types contributing to the correlator
 \re{freecorr} at tree level. The lines denote free scalar propagators $\vev{ \phi (x_i) \phi (x_j)}$.
 In the light-cone limit $x_{i,i+1}^2\to 0$ the leading
 contribution comes from diagram (a), while that of diagram
 (b) is suppressed by the factor $x_{34}^2 x_{1n}^2/(x_{3n}^2x_{14}^2)$.
 \label{ttree}}
 \end{figure}

Let us start with a simple example. {Consider the $n-$point correlation function $\langle {\cal O}(x_1) \cdots
{\cal O}(x_n) \rangle$ of the operators ${\cal O}(x) = {\rm Tr}[\phi^2(x)] $
 in a free theory with  $\phi(x)$ being a real scalar
field.}  When we approach the limit $x_{i,i+1}^2 \to 0 $, the
most singular part of the  connected
correlator   goes as
\be \label{freecorr}
\langle {\cal O}(x_1) \cdots
{\cal O}(x_n) \rangle \propto { N_c^2 \over \prod_{i=1}^n x_{i,i+1}^2 }\,.
\ee
This is certainly one of the terms { contributing to} the free correlator. There are other terms which are less singular
where some of the contractions involve space-like separated points, see Figure \ref{ttree}.

Let us now consider the same operators in an interacting theory which is dimensionally regularized.
We expect that in the limit $x_{i,i+1}^2 \to 0$ the leading singularity in the
correlator is the same as in the free theory, Eq.~\nref{freecorr}. The reason is that
 we are taking the limit with the regularization scale $\mu^2$ kept fixed,  $x_{i,i+1}^2 \mu^2 \to 0$,
 so that when the distances are becoming zero, we are exploring the UV structure of the theory which
 is free in the regularized theory. Here we are using that dimensional regularization preserves Lorentz
 invariance, even for distances smaller than the regulator scale.

 Naively one would think that the limit is then identical to \nref{freecorr}. However, the fact that the
 $\phi-$particles are color charged implies that they source a color electric field. The electric field is not
 modifying the singular behavior of the correlator, but it leads to a finite contribution 
 \be\label{2.2}
 \lim_{x_{i,i+1}^2\to 0 } { G_n \over G_n^{\rm tree} } = W^{{\rm adj }}_\epsilon [C_n] = \frac1{N_c^2-1}\vev{0|\Tr_{\rm adj}  {\rm P}\exp \lr{ig\oint_{C_n} dx\cdot A(x) }|0}\,,
\ee
where $W^{{\rm adj }}_\epsilon [C_n] $ is the Wilson loop in the adjoint representation in
the dimensionally regularized theory.  The contour $C_n$ is a polygonal contour where all sides are null and the cusps are located at the limiting
 positions of the operator insertions.
This formula would also be valid in a theory that is UV free such as $(2+1)-$dimensional Yang Mills.\footnote{Even
though QCD is asymptotically free, the coupling does not decrease fast enough to render these Wilson loops finite
 in four dimensions. }

In the planar approximation the adjoint Wilson loop is equal to the product of a Wilson loop in the
fundamental and one in the anti-fundamental. In a charge conjugation { invariant}   theory the last two
are equal { to each other} and we can write $W^{\rm adj}[C_n] = ( W[C_n])^2 $.

\subsection{A more rigorous explanation }\label{cflc}

In this section { we argue}  that the correlation functions of protected operators
in {any conformal field theory
have a universal  behavior in the light-cone limit $x_{i,i+1}^2\to 0$, similar to that of a polygonal light-like Wilson loop.

At tree level, the $n-$point correlation function  $G_n^{\rm tree}$ of  operators of the schematic form  $\Tr[\phi^2(x)]$
reduces to a sum of products of $n$ free scalar propagators.
The corresponding Feynman diagrams can be separated into
connected and disconnected ones. The contribution of the connected graphs
has the form
\begin{align}\label{D-tree}
G_n^{\rm tree} = N_c^2 \sum_{\{i_1,\ldots,i_n\} }S(x_{i_1,i_2})S(x_{i_2,i_3})\ldots S(x_{i_n,i_1}) \,,
\end{align}
where $x_{ij}=x_i-x_j$ and $S(x) = 1/( 4\pi^{2}\, x^2)$ is the free scalar propagator. Here the sum runs over all non-cyclic permutations of the indices $i_1,\ldots,i_n$. In the light-cone limit $x_{i,i+1}^2\to 0$ (with $x_{i+n}\equiv x_i$) the dominant
contribution to \re{D-tree} comes from a single term with cyclic ordering of the indices:
\begin{align}\label{D-tree-lc}
G_n^{\rm tree} \stackrel{x_{i,i+1}^2 \to 0}{\longrightarrow} N_c^2\,  S(x_{12})S(x_{23})\ldots S(x_{n1})
=\frac{(2\pi)^{-2n}N_c^2}{x_{12}^2x_{23}^2\ldots x_{n1}^2} \,.
\end{align}
Here we have assumed that the operators are such that the
contraction leads to a non-vanishing result. If we consider bilinear operators in the
${\cal N}=4$ super Yang Mills theory, this statement means that we
are choosing Konishi operators or the half-BPS operators in the ${\bf 20'}$ in
the appropriate fashion. At loop level, we have to take into account
that, propagating between the points $x_i$ and $x_{i+1}$, the scalar
field can interact with gluons, gluinos and with other scalars.
Examples of the corresponding diagrams are shown in
Fig.~\ref{fig:S}. In what follows we shall compute their
contribution in two steps. We will first discard the interaction
with gluinos and scalars and resum the diagrams shown in
Fig.~\ref{fig:S}(a) over all possible gluons exchanges. Then, we
will take into consideration the remaining interaction vertices
shown in Fig.~\ref{fig:S}(b)--(d) and argue that their contribution
does not affect the leading asymptotic behavior of the correlator
for $x_{i,i+1}^2\to 0$.
\begin{figure}[h]
\psfrag{(a)}[cc][cc]{(a)}
\psfrag{(b)}[cc][cc]{(b)}
\psfrag{(c)}[cc][cc]{(c)}
\psfrag{(d)}[cc][cc]{(d)}
\psfrag{xi}[cc][cc]{$x_i$}
\psfrag{xj}[cc][cc]{$x_{i+1}$}
\centerline{ \includegraphics[width=\textwidth]{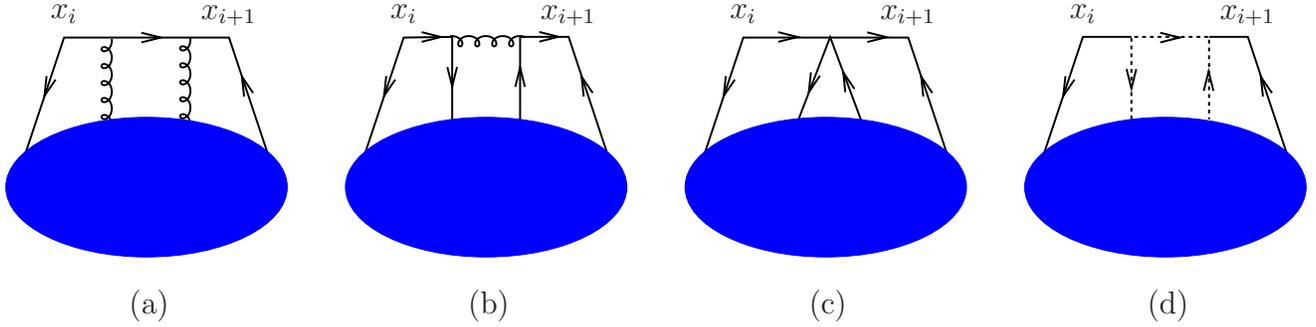} }
 \caption{\small {Different types of diagrams contributing to the
correlation function $G_n$. Solid, wavy and dashed lines denote scalars,
gluons
and gluinos, respectively. The vertex with coordinate $x_i$
represents the operator $\cO(x_i)$. The shadowed blobs stand for the
  rest of the diagram involving the remaining operators. }} \label{fig:S}
 \end{figure}

It is easy to see that the net effect of the interaction of the scalar particle, propagating
between the points $x_i$ and $x_{i+1}$, with the gauge field amounts to
replacing the free scalar propagator, $S(x_{i,i+1})$, by the propagator in a
background gluon field,  $S(x_i,x_{i+1};A)$,
\begin{align}\label{G-loop}
G_n  \stackrel{x_{i,i+1}^2 \to 0}{\longrightarrow} \vev{0| \Tr [ {S}(x_1,x_2;A) S(x_2,x_3;A)\ldots S(x_n,x_1;A)]|0}\,,
\end{align}
where the expectation value is taken with respect to the $\cN=4$ SYM action \re{lagrangian}. By definition, the scalar propagator
in the external field, $S(x,y;A)$, satisfies the equation
\begin{align}\label{D-eq}
i D^\mu D_{\mu}\, S  (x,y;A) = \delta^{(4-2\ep)}(x-y)  \,,
\end{align}
where $D^\mu = \partial_x^\mu - ig[A^\mu(x),\ ]$. Here we have introduced
dimensional regularization in order to deal with the short-distances singularities
of the correlator for $x_{i,i+1}^2\to 0$.

According to \re{G-loop}, the asymptotic behavior of the correlator
on the light-cone is related to the behavior of the scalar propagator $S(x_i,x_{i+1};A)$ in the limit $x_{i,i+1}^2\to 0$. In this limit, it is convenient to look for a solution to \re{D-eq} of the form
\begin{align}\label{ansatz}
S(x_i,x_{i+1};A) =S^{\rm tree}(x_{i,i+1}) P\exp\lr{ig\int^{x_{i+1}}_{x_i} dz\cdot   A(z)} G(x_i,x_{i+1};A)\,,
\end{align}
where we have singled out the free scalar propagator
$S^{\rm tree}(x_{i,i+1}) =  (-x_{k,k+1}^2)^{-1+\ep}  {\Gamma(1-\ep)}/{(4\pi^{2-\ep})}$
and have factorized the dependence on the gauge field into a path-ordered exponential running along the line connecting the points $x_i$ and $x_{i+1}$, and yet another (matrix-valued) function $G(x_i,x_{i+1};A)$ to be determined below. Notice that the path-ordered exponential in \re{ansatz} is defined in the {\em adjoint} representation of the gauge group, $[ A_\mu(z)]^{ab} = if^{abc} A^c_\mu(z)$.
It transforms under gauge transformations in the same way as the scalar propagator,  $S(x_i,x_{i+1};A^\Omega) = \Omega^{-1}(x_{i})   S(x_i,x_{i+1};A) \Omega(x_{i+1})$. Then the second factor in the right-hand side of \p{ansatz} transforms as follows
\begin{align}\label{S-gauge}
G(x_i,x_{i+1};A^\Omega) = \Omega^{-1}(x_{i+1}) G(x_i,x_{i+1};A) \Omega(x_{i+1})\,.
\end{align}
For our purposes, we need to know the expansion of $G(x_i,x_{i+1};A)$ for $x_{i,i+1}^2\to 0$.

The  expansion of the scalar propagator in a background gauge field
crucially depends on the hierarchy between the two scales,
$1/x_{i,i+1}^2$ and $\mu^2$. The former  defines the energy carried
by the scalar field, while the latter sets up an ultraviolet cutoff
on the momentum carried by the gauge field. In this section, we are
interested in the limit $x_{i,i+1}^2 \mu^2 \ll 1$. It corresponds to
the situation where the gluon momentum is much smaller than the
energy of the scalar particle. In other words, the propagator
\re{ansatz} describes the scattering of  an infinitely fast scalar
(color charged) particle off a slowly varying background field
$A_\mu(x)$.
Such a scattering only slightly modifies the free scalar propagator
by inducing the eikonal phase given by
$P\exp\lr{ig\int^{x_{i+1}}_{x_i} dz\cdot  A(z)}$.
%
%
In terms of \re{ansatz} this corresponds to
\begin{align}\label{one}
G(x_i,x_{i+1};A)\to 1\,,\qquad  \text{as\ \  $x_{i,i+1}^2 \mu^2 \to 0$}\,.
\end{align}

The same result can be derived from the conventional operator product expansion (OPE). Let us return to \re{ansatz} and examine the asymptotic behavior of the scalar propagator at short distances, $x_{i,i+1}^2\to 0$. It is convenient to choose $x_i=x$, $x_{i+1}=0$ and study $G(x,0;A)$ for $x^2\to 0$. In this limit, we can apply the OPE and expand $G(x,0;A)$ over an (infinite) set of local operators. 
Then, taking into account \re{S-gauge}, we observe that  the expansion of  $G(x,0;A)$ at small $x^2$ should run over local operators $\mathcal{O}^{\mu_1\ldots \mu_N}(0)$ built from the gauge field strength $F_{\mu\nu}$ and the covariant derivatives $D_\mu$  at the origin,
\begin{align}\label{OPE}
G(x,0;A) =  \sum_{N,\Delta} (x^2)^{\Delta} C_{\Delta,N}(x^2\mu^2)\ x_{\mu_1}\ldots  x_{\mu_N}  \mathcal{O}_\Delta^{\mu_1\ldots \mu_N}(0)\,,
\end{align}
where the expansion goes over local operators carrying Lorentz spin $N$ and canonical dimension  $\ell_{ \mathcal{O}}$. Here $C_{\Delta,N}(x^2\mu^2)$ are dimensionless coefficient functions and the Wilson operators $\mathcal{O}^{\mu_1\ldots \mu_N}(0)$ are normalized at the scale $\mu^2$. For $x^2\to 0$ the contribution of local operators to the right-hand side of \re{OPE}
is suppressed as $(x^2)^\Delta$. It follows from the dimensional counting that the exponent $\Delta$ is determined by the twist of the operator $\tau=\ell_{ \mathcal{O}}-N$
\begin{align}\label{twist}
2\Delta+N = \ell_{ \mathcal{O}} \qquad \to\qquad  \Delta=\tau/2\,,
\end{align}
with $\tau$ taking non-negative integer values.
If the twist of $\mathcal{O}_\Delta^{\mu_1\ldots \mu_N}$ is non-vanishing, $\tau = 1,2,\ldots$, we have $\Delta>0$; if instead $\tau=0$, which corresponds to  the identity operator, we find $\Delta = 0$. So,  the leading contribution to \re{OPE} in the limit $x^2\to 0$ comes from the identity operator.\footnote{ The explicit expression for the subleading contribution of the  twist-two operators to \re{OPE} can be found in Appendix~\ref{App-scalar}. }

Notice that the relation \re{OPE} holds for arbitrary $x^2$ and $\mu^2$. The physical meaning of \re{OPE} is that the OPE separates the contributions
from large and short distances, compared to $1/\mu^2$, into local operators and coefficient functions, respectively. Let us examine \re{OPE} in two different limits:
$x^2\mu^2 \ll 1$ and $x^2\mu^2 \gg 1$. In dimensional regularization, the perturbative expansion of the coefficient functions $C_{\Delta,N}(x^2\mu^2)$ goes in powers of $g^2 (x^2\mu^2)^\ep$. As a consequence, for $x^2\mu^2 \to 0$ the {loop corrections to the} coefficient function  $C_{\Delta,N}(x^2\mu^2)$ vanish order by order in the coupling constant and we recover the expected result \re{one}. At the same time, for $x^2\mu^2 \gg 1$ the coefficient function is different from unity and needs to be taken into account.

So far we considered the contribution of the diagrams shown in Fig.~\ref{fig:S}(a). Let us now examine the remaining diagrams in Figs.~\ref{fig:S}(b)--(d). In a close analogy with the previous case, we can interpret them as contributing to the propagator of a scalar particle in the background of the other scalars and gluinos. As before, the short-distance behavior of the propagator
can be studied using the OPE \re{OPE}. The net effect of the diagrams  in Figs.~\ref{fig:S}(b)--(d) is to enlarge the set of local operators contributing
to the right-hand side of \re{OPE}. Namely, the resulting local operators will involve
additional pairs of scalars and gluinos. Most importantly, the twist of such operators is greater than two. Consequently, their contribution to \re{OPE} is suppressed as $x^2\to 0$ and, therefore, it can be safely discarded.

Finally, we combine together the relations \re{ansatz} and \re{one} and
obtain the leading asymptotic behavior of the propagator on the light cone $x_{i,i+1}^2\mu^2\to 0$ as   \cite{Gross:1971wn,IKR85}
\begin{align}\label{S-as}
S(x_i,x_{i+1};A) \to S^{\rm tree}(x_{i,i+1}) P\exp\lr{ig\int^{x_{i+1}}_{x_i} dz\cdot   A(z)}  \,.
\end{align}
This relation has a transparent physical interpretation. In the first-quantized
picture, the propagator $S(x_i,x_{i+1};A)$ describes the transition amplitude
for a charged massless particle to go from point $x_i$ to $x_{i+1}$. As such, it is
given by the sum over all paths $C_{x_i,x_{i+1}}$ connecting these two points. The interaction
of the particle with the external gauge field brings in an additional weight factor given by the path-ordered exponential evaluated along $C_{x_i,x_{i+1}}$. In the limit $x_{i,i+1}^2\mu^2 \to 0$, corresponding to the propagation of an infinitely fast particle along the light cone, the path
integral is dominated by the saddle point contribution. The latter is just
the classical trajectory of the particle, that is, the straight line connecting points $x_i$ and $x_{i+1}$.

Let us now replace the scalar propagators in \re{G-loop} by their leading asymptotic behavior \re{S-as}. We find that the product of free propagators reproduces
the correlator at tree level, Eq.~\re{D-tree-lc}, whereas the
path-ordered exponentials are combined into a single factor,
\begin{align}\label{duality}
G_n\  {\to} \  G^{\rm tree}_n \,W^{\rm adj}[C_n]\,,
\end{align}
where $G^{\rm tree}_n$ is the tree-level correlator and
$W^{\rm adj}[C_n]$ is a Wilson loop in the {\em adjoint} representation,  evaluated
along the light-like polygon $C_n=[x_1,x_2]\cup [x_2,x_3]\cup\ldots\cup [x_n,x_1]$:
\begin{align}\label{W-adj}
W^{\rm adj}[C_n] =\frac1{N_c^2-1}\, \vev{0|\Tr_{\rm adj}\left[ P\exp\lr{ig \int_{C_n} dz\cdot   A(z)}\right] |0}\,.
\end{align}
{Here the overall normalization factor
$1/(N_c^2-1)$ is inserted  in order for $W^{\rm adj}[C_n]$ to be 1 at the lowest order in the coupling constant.}
Since the tree-level expression  $G^{\rm tree}_n$ is not well defined
for $x_{i,i+1}^2=0$, we can rewrite \re{duality} in the
form
\begin{align}\label{intheadj}
\lim_{x_{i,i+1}^2\to 0}\lr{ \frac{G_n}{G^{\rm tree}_n}} = W^{\rm adj}[C_n]\,.
\end{align}
In Sect.~\ref{explicit} we check  the validity of this
 relation
 by an explicit one-loop calculation in $\mathcal{N}=4$ SYM theory.

Notice that the Wilson loop $W^{\rm adj}[C_n]$ contains UV divergences
due to the cusps on the contour.   This does not contradict, however, the UV finiteness of the correlator of protected operators $G_n$. It is
only in the light-cone limit, $x_{i,i+1}^2\to 0$, that the correlator
develops new, light-cone singularities.  According to \re{intheadj}, they are in one-to-one correspondence with the ultraviolet divergences of the light-like Wilson loop.

 We wish to emphasize that the Wilson loop in the right-hand side of
\re{intheadj} is defined in the adjoint representation of the color group $SU(N_c)$. This
has to do with the fact that the $\cN=4$ SYM scalars {belong to this representation.}  The relation  \re{intheadj} can be further
simplified in the multi-color limit by using the well-known property
of the Wilson loops {mentioned at the end of Sect.~2.1},
\begin{align}\label{W-AF}
W^{\rm adj}[C_n] =  \big(W[C_n]\big)^2 + O(1/N_c^2)\,,
\end{align}
where $W[C_n]$ is defined in the {\em fundamental} representation of $SU(N_c)$, see Eq.~\re{Wl}. The relation \re{W-AF} becomes very useful when comparing correlators with planar gluon MHV scattering amplitudes, since the latter are related to Wilson loops $W[C_n]$ in the fundamental representation.

\section{Explicit computation in ${\cal N}=4$ super-Yang-Mills }
\label{explicit}

In this section we perform some explicit computations in ${\cal N}=4$ super-Yang-Mills to confirm the
above picture. The reader who is interested only in general statements can jump to the next section.

In the $\cN=4$ SYM theory there are two types of gauge invariant operators, protected and non-protected. The former are not renormalized and thus have a fixed conformal dimension, equal to their canonical dimension.  The simplest example is the bilinear gauge invariant operator made of the six real scalars in the vector multiplet, $\phi_{AB} = -\phi_{BA}  = \ft12  \ep_{ABCD}\bar\phi^{CD}$, where $A,B=1,2,3,4$ are indices of the fundamental irrep of the R symmetry group $SU(4)$. The bilinear
\begin{equation}\label{bil}
    \cO_{ABCD} = {\rm Tr}(\phi_{AB} \phi_{CD}) -  \frac1{12}\ep_{ABCD} {\rm Tr}(\bar\phi^{EF} \phi_{EF})
\end{equation}
belongs to the irrep $\bf 20'$ of $SU(4)$. This scalar operator is the superconformal primary state of a whole tower of protected operators containing, among others, the stress tensor of the theory. Such ``short'', or half-BPS operators (also known as CPO) have been extensively studied (see, e.g., \cite{Andrianopoli:1998jh}) in the context of the AdS/CFT correspondence \cite{Maldacena:1997re}. The best known example of an unprotected (i.e., having an anomalous dimension) operator is the so-called Konishi operator
\begin{equation}\label{Ko}
{\cal K} = \mathrm{Tr}( \bar \phi^{AB} \phi_{AB} ) \, .
\end{equation}

Here we present two examples which illustrate how the correlator becomes a Wilson loop in the light-cone limit. The first is a correlator involving only protected operators, in the second we replace some of these by Konishi operators. In the first case, it is sufficient to consider only certain projections  of $\bf 20'$, namely
\begin{align}\label{2f}
\cO = {\rm Tr}(\phi_{12} \phi_{12})\,,\qquad\qquad
\tilde\cO = {\rm Tr}(\bar\phi^{12} \bar\phi^{12})\,,
\end{align}
where $\cO$ is the  highest-weight state and $\tilde\cO$ is the conjugate lowest-weight state.
We want to evaluate the correlator of $n=2m$ operators of the form\footnote{For $n=2m+1$ we can add, e.g., one real  operator $\hat\cO= {\rm Tr}(\bar\phi^{12} \phi_{12}) -  \frac{1}{12}
{\rm Tr}(\bar\phi^{EF} \phi_{EF})$.}
\begin{equation}\label{defco}
    G_n = \vev{\cO(x_1) \tilde\cO(x_2)\ldots \cO(x_{n-1})\tilde\cO(x_n)}\,.
\end{equation}
Such correlators are finite  and conformally covariant, because the operators $\cO$ are not renormalized. Moreover, it is known that the two- and three-point correlators are themselves protected, i.e. they do not receive quantum corrections beyond tree level \cite{Penati:1999ba,Penati:2000zv,D'Hoker:1998tz,Lee:1998bxa,Howe:1998zi}. Here we are interested in the loop corrections to $G_n$, therefore we consider the cases $n\geq 4$. Then, the loop corrections to $G_n$ are given by conformally invariant space-time integrals.\footnote{For an extensive study of such correlators in the case $n=4$, at one and two loops, see, e.g., \cite{Eden:2000mv,Bianchi:2000hn} and references therein.}

Next, we wish to take the limit when the neighboring points become light-like separated,
\begin{equation}\label{lim}
    x^2_{i,i+1} \ \to \ 0\,, \qquad i=1,\ldots,n
\end{equation}
(assuming that $x_{n+1} \equiv x_1$).
This limit is singular for two reasons. Firstly, the correlator develops pole singularities, as can be seen already from the (connected, planar) tree-level approximation
\begin{equation}\label{tre'}
    G_n^{\rm tree} =   \frac{(2\pi)^{-2n}N^2_c}{x^2_{12} x^2_{23} \ldots x^2_{n1}} + \mbox{subleading terms}\,.
\end{equation}
By ``subleading" we mean terms corresponding to different Wick contractions of the fields $\phi$ which are less singular in the limit \p{lim} (see an illustration in Fig.~\ref{ttree}). This   can be remedied by considering the ratio
\begin{equation}\label{ratonc}
    \lim_{x^2_{i,i+1}\to 0}G_n/G^{\rm tree}_n\,.
\end{equation}

Secondly, the loop integrals develop short-distance (ultraviolet) logarithmic divergences, which can be regularized by computing the correlator in $D=4-2\ep$ (with $\ep>0$) dimensions. Our aim in this section is to show the relation
\begin{equation}\label{coWL}
   \lim_{x^2_{i,i+1}\to 0}G_n/G^{\rm tree}_n = (W[C_n])^2\,,
\end{equation}
where $W[C_n]$ is the expectation value of a Wilson loop evaluated on a polygonal light-like contour $C_n$ with $n$ cusps at the points $x_i$,
\begin{align}\label{Wl}
 W[C_n]=\frac1{N_c}\vev{0|\tr {\rm P}\exp \lr{ig\oint_{C_n} dx\cdot A(x) }|0}\,,
\end{align}
and computed in the {same UV regularization scheme} as the correlator.
Here $W[C_n]$ is defined in the fundamental
representation of $SU(N_c)$, with $A_\mu(x) = A_\mu^a t^a$
and $t^a$ being the generators of the fundamental representation.
{To lowest order in the coupling, the Wilson loop takes the form
\begin{align}\label{lo}
W[C_n] =  1+\frac{1}{4}(ig)^2N_c \oint_{C_n} dx^\mu \oint_{C_n} dy^\nu \,
 D_{\mu\nu}(x-y)  + O(g^4)\,,
\end{align}
where $D_{\mu\nu}(x)$ is the free gluon propagator.

The expression in the right-hand side of \re{coWL} can be rewritten as
a Wilson loop $W^{\rm adj}[C_n]$ in the {\em adjoint} representation of $SU(N_c)$, see Eqs.~\re{W-adj} and \p{W-AF}. The reason why we prefer to formulate the relation \p{coWL} in terms of the Wilson loops $W[C_n]$ is that they are known to match planar gluon MHV scattering amplitudes \cite{am07,Drummond:2007aua,Brandhuber:2007yx}. }

\label{Lip}

 The correlator \p{defco} is given by the path integral
\begin{equation}\label{corcomp}
   G_n = \int {\cal D}\Phi\ e^{i\int d^4x {L}(x)}\ \cO(x_1) \ldots \tilde\cO(x_n)\,,
\end{equation}
where $\Phi$ denotes all the fields of the theory. The Lagrangian  of $\cN=4$ SYM has the form
\begin{eqnarray}
{L}_{{\cal N} = 4} &=& {\rm Tr} \, \left\{\half
F_{\mu\nu} F^{\mu\nu} + \half \left( { D}_\mu \bar\phi^{AB} \right)
\left( {D}^\mu \phi_{AB} \right) + \frac{1}{8} g^2 [\bar\phi^{AB},
\bar\phi^{CD}] [\phi_{AB}, \phi_{CD}] \right.
\nonumber\\
&+& \left. \!\!\! 2 i \bar\lambda_{\dot\alpha A} \sigma^{\dot\alpha
\beta}_\mu { D}^\mu \lambda^A_\beta - \sqrt{2} g \lambda^{\alpha
A} [\phi_{AB}, \lambda_\alpha^B] + \sqrt{2} g
\bar\lambda_{\dot\alpha A} [\bar\phi^{AB}, \bar\lambda^{\dot\alpha}_B]
\right\}\,,  \label{lagrangian}
\end{eqnarray}
with ${D}^\mu  = \pa^\mu  - ig [A^\mu, \ ]$.

Instead of computing the loop corrections to this correlator directly, we prefer to use the well-known procedure of differentiating $G_n$ with respect to the coupling $g$. Before doing this, we rescale the gauge field,
\begin{equation}\label{resc}
    A_\mu \ \to \ g^{-1} A_\mu\,.
\end{equation}
As a result, the gluon Lagrangian picks an overall factor, $(1/2g^2)\, {\rm Tr}(F_{\mu\nu} F^{\mu\nu})^2$, and the coupling drops out of the covariant derivatives, ${D}^\mu  = \pa^\mu  - i [A^\mu, \ ]$. Then the derivative of $G_n$ with respect to  the coupling
produces
\begin{eqnarray}
  g^2\frac{\pa}{\pa g^2} G_n( x_1, \ldots, x_n) &= & -i\int {\cal D}\Phi\ e^{i\int d^4 x {L}(x)}\ \int d^4 x_0 {L}'(x_0) \cO(x_1) \ldots \tilde\cO(x_n) \nn \\
  &\equiv& -i\int d^4 x_0\, {\cal G}_{n+1}(x_0, x_1, \ldots, x_n)\,. \label{derG}
\end{eqnarray}
Here
\begin{equation}\label{n+1co}
    {\cal G}_{n+1} = \vev{{L}'(x_0) \cO(x_1) \ldots \tilde\cO(x_n)}
\end{equation}
is  a new, $(n+1)$-point correlator obtained by inserting the derivative of the  Lagrangian
\begin{equation}\label{L'}
    {L}' = {\rm Tr} \, \left\{- \frac1{2g^2} F_{\mu\nu} F^{\mu\nu} + \frac{g^2}{8}  [\bar\phi^{AB},
\bar\phi^{CD}] [\phi_{AB}, \phi_{CD}] - \frac{g}{\sqrt{2}} \lambda^{\alpha
A} [\phi_{AB}, \lambda_\alpha^B] + \frac{g}{\sqrt{2} }
\bar\lambda_{\dot\alpha A} [\bar\phi^{AB}, \bar\lambda^{\dot\alpha}_B]
\right\}
\end{equation}
into the original correlator. This new correlator is then integrated in \p{derG} over the  insertion point $x_0$, which generates the loop corrections to $G_n$. The advantage of this procedure is that we gain one perturbative order. In particular, the one-loop correction $G_n^{(1)}$ is determined by the tree- (or Born) level correlator ${\cal G}_{n+1}^{(0)}$.

\subsection{Computing the correlator with a single insertion}

We need to evaluate  ${\cal G}_{n+1}$ at tree level. The rescaling of the gauge field \p{resc} modifies the gluon propagator by a factor of $g^{2}$, while the vertices involving the gauge field now appear without the coupling.  In addition, the various terms in the Lagrangian insertion in \p{n+1co} have their own factors of $g$. Putting all these factors together, we confirm that the tree-level correlator   ${\cal G}_{n+1}^{(0)}$ is of order $g^2$, in accordance with \p{derG}. Inspecting the terms in the inserted Lagrangian ${L}'$, we see that there exist two types of Feynman graphs at this order shown in Fig.~\ref{2loopgraphs'}.\footnote{Notice that the graphs are drawn with a polygonal matter frame. Graphs based on the ``zigzag" configurations like in Fig.~\ref{ttree}(b) are suppressed in the light-cone limit \p{lim}, after dividing out the leading singularity of the tree-level correlator \p{tre'}. \label{zigf}}

\medskip

\begin{figure}[h!]
\psfrag{dots}[cc][cc]{ }
\psfrag{0}[cc][cc]{$x_0$}
\psfrag{1}[cc][cc]{$x_{l+1}$}
\psfrag{2}[cc][cc]{ }
\psfrag{3}[cc][cc]{$x_k$}
\psfrag{4}[cc][cc]{$x_{k+1}$}
\psfrag{n-1}[cc][cc]{ }
\psfrag{n}[cc][cc]{$x_l$}
\psfrag{(a)}[cc][cc]{(a)}
\psfrag{(b)}[cc][cc]{(b)}
\psfrag{(c)}[cc][cc]{(c)}
\psfrag{=}[cc][cc]{=}
\psfrag{+}[cc][cc]{+}
%
\centerline{ \includegraphics[width=\textwidth]{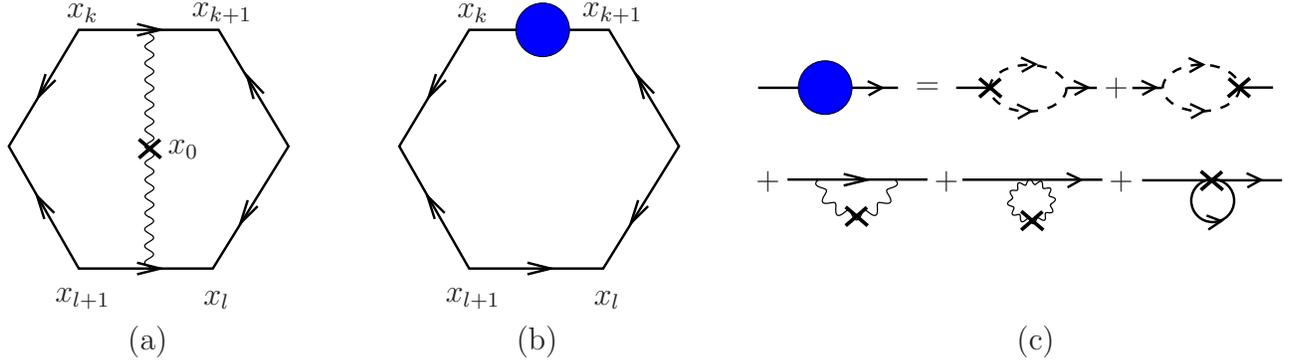} }
 \caption{\small The leading contribution to the correlator \re{n+1co} in the light-cone limit. Solid, wavy and dashed lines denote complex scalars, gluons and gluinos, correspondingly. The cross denotes the insertion of the derivative of the  Lagrangian with respect to the coupling constant ${L}'(x_0)$, Eq.~\re{L'}.
The big blob in diagram (b) denotes the corrections to the scalar propagator shown in
(c). \label{2loopgraphs'} }
 \end{figure}

After integration over the insertion point, the graphs in Fig.~\ref{2loopgraphs'}(b), (c) give the correction to the two-point function (propagator) $\vev{\bar\phi(x_k) \phi(x_{k+1})}$. As explained in Sect.~\ref{cflc}, we are interested in the particular limit $x^2_{i,i+1} \to 0$, by keeping $\mu^2 x^2_{i,i+1} \ll 1$. In this limit, on dimensional grounds,
\begin{equation}\label{dimgr}
    \lim_{x^2_{k,k+1} \to 0} \frac{\vev{\bar\phi(x_k) \phi(x_{k+1})}}{\vev{\bar\phi(x_k) \phi(x_{k+1})}^{\rm tree}} = \lim_{\mu^2 x^2_{k,k+1} \to 0} [1+ g^2 (\mu^2 x^2_{k,k+1})^{\ep} C(\ep) + O(g^4)] = 1\,.
\end{equation}
So, these graphs do not contribute on the light cone.

The  non-trivial contribution comes from the graph in Fig.~\ref{2loopgraphs'}(a). It contains the gauge kinetic term $F^2$ at the insertion point.\footnote{The quartic scalar term $g^2{\rm Tr}([\bar\phi^{AB},
\bar\phi^{CD}] [\phi_{AB}, \phi_{CD}])$ from \p{L'} does not appear due to the special choice of the external scalars.} Summing up all such graphs, we get the following  ratio
\begin{equation}\label{grapha}
    G_n/G^{\rm tree}_n = -i g^2 N_c \sum_{k\neq l=1}^n
    \frac{\int d^D x_0 \ T^{\mu\nu}(x_k, x_0, x_{k+1})\, T_{\mu\nu}(x_l, x_0, x_{l+1})}{S(x_{k,k+1}) S(x_{l,l+1})} +O(g^4) \,,
%
\end{equation}
where $S(x)$ is the free scalar propagator in $D=4-2\ep$ dimensions,
\begin{align}\label{pro}
{\vev{\bar\phi_a (x_k) \phi_b(x_{k+1})}^{\rm tree}}= \delta_{ab} S(x_{k,k+1}) =\delta_{ab} \frac{\Gamma(1-\ep)}{4\pi^{2-\ep}} (-x_{k,k+1}^2+i0)^{-1+\ep}\,,
\end{align}
and $T^{\mu\nu}(x_k, x_0, x_{k+1})$ is defined by the three-point correlator
\begin{equation}\label{Tvert}
    \vev{\bar\phi_a(x_k) F_b^{\mu\nu}(x_0) \phi_c(x_{k+1})}= g f_{abc} T^{\mu\nu}(x_k, x_0, x_{k+1})\,.
\end{equation}
To lowest order in the coupling it
has the form (we use Feynman gauge for the gluon propagator)
\begin{align}
T^{\mu\nu}(x_k, x_0, x_{k+1}) &=   \int d^{4-2\ep}x_{0'}  \left(
   S(x_{k,0'}) {\overset{\leftrightarrow}{\pa}}_{x_{0'}}^\lambda  S(x_{k+1,0'})\right) \left(\delta_\lambda^{[\mu} \pa_{x_{0}}^{ \nu]} S(x_{00'})\right) \nn\\
   &=   \kappa_\ep^3\,
   \partial_k^{[\mu}\partial_{k+1}^{\nu]}\int \frac{d^{4-2\ep} x_{0'}}{(-x_{k,0'}^2 x_{k+1,0'}^2 x_{00'}^2)^{1-\epsilon}} 
   \equiv
    \kappa_\ep^3\,  x_{k,0}^{[\mu}x_{k+1,0}^{\nu]}\ I_\epsilon (x_{k0}, x_{k+1,0})\,,  \label{vertdi}
\end{align}
where $\kappa_\ep = {\Gamma(1-\ep)}/{(4\pi^{2-\ep})}$ and
${\overset{\leftrightarrow}{\pa}}_{x} =\ft12( {\overset{\rightarrow}{\pa}}_{x}-{\overset{\leftarrow}{\pa}}_{x} )$.

If we keep the points $x_k$ and $x_{k+1}$ separated, $x_{k,k+1}^2\neq 0$, the integral in
\p{vertdi} converges and we can set $D=4$, i.e., $\ep=0$. In this case the scalar integral $I_0$ becomes conformally covariant and can easily be evaluated (see, e.g., \cite{Eden:1998hh} and also Appendix~\ref{apA}),
\begin{equation}\label{I0}
    I_{\ep=0}(x_{k0}, x_{k+1,0})= -\frac{4i\pi^2}{x^2_{k,k+1}x^2_{k,0} x^2_{k+1,0}}\,.
\end{equation}
In the limit  $x^2_{k,k+1} \to 0$ the integral diverges and we have to stay in $D=4-2\ep$ dimensions. In Appendix~\ref{apA} we show that for  $x_{k,k+1}^2\to 0$
\begin{align}\label{intI}
I_\epsilon (x_{k0},x_{k+1,0})&\  {\to} - i  (1-\ep)\kappa_\ep^{-1}\,  (-x_{k,k+1}^2)^{-1+\epsilon}\,
 \int_0^1 ds\, [-(x_{k,0} s + x_{k+1,0} \bar s)^2]^{-2+\epsilon} \,,
\end{align}
with $\bar s=1-s$.
We see that the asymptotic singular behavior of the vertex correction \p{vertdi}
for $x_{k,k+1}^2\to 0$ is the same as that of the free propagator \re{pro}.

Next, replacing  $I_\epsilon(x)$ in \p{vertdi} by its asymptotic form \p{intI} and substituting the vertex corrections in \p{grapha} yields
\begin{align}\label{L}
 \lim_{x^2_{i,i+1}\to 0} G_n/G^{\rm tree}_n = -2 i g^2 N_c [  (1-\ep) \kappa_\ep]^2
   \sum_{k\neq l=1}^n \int_0^1 ds_k ds_l \int d^{4-2\ep}x_0\,
\frac{ x_{a0}^{[\mu} x_{k,k+1}^{\nu]} (x_{b0})_{\mu} (x_{l,l+1})_{\nu}}{[x_{a0}^2x_{b0}^2]^{2-\epsilon}}\,,
\end{align}
with the notation
\begin{align}
x_a= x_{k} s_k + x_{k+1}\bar s_k\,, 
\qquad
x_b  =x_{l} s_l + x_{l+1}\bar s_l\,.
\end{align}
Up to a normalization factor the above integral can be
rewritten as
\begin{align}
-\frac{i}2 \kappa_\ep^2   x_{k,k+1}^\mu x_{l,l+1}^\nu \left[ (\partial_{a})_\mu (\partial_{a})_\nu - g_{\mu\nu} \square_{a}\right] \int \frac{d^{4-2\ep}x_0\,  }{[x_{0a}^2 x_{0b}^2]^{1-\epsilon}} = -\frac12 x_{k,k+1}^\mu  x_{l,l+1}^\nu D_{\mu\nu}(x_{ab})\,,
\end{align}
where $D_{\mu\nu}(x_{ab})$ is the gluon propagator in the {\it Landau gauge}
\begin{align}
D_{\mu\nu}(x) = -i\int \frac{d^{4-2\ep} k}{k^2}\e^{ikx}\, \lr{g^{\mu\nu}-\frac{k^\mu k^\nu}{k^2}}\,.
\end{align}
Returning to \re{L}, we conclude that
\begin{align}
\lim_{x^2_{i,i+1}\to 0} G_n/G^{\rm tree}_n &=  \frac12 (ig)^2N_c \sum_{k\neq l} \int_0^1 ds_k \int_0^1 ds_l\,
 x_{k,k+1}^\mu  x_{l,l+1}^\nu
D_{\mu\nu}(x_{ab})  + O(g^4) \notag\\
& = 2\log W[C_n]\,,
\end{align}
which is just the one-loop expression for the light-like Wilson loop \p{lo} in the fundamental representation of the $SU(N_c)$ gauge group, calculated  {in the Landau gauge}.
Since the Wilson loop is gauge invariant,  we can claim that this result is equivalent to the earlier Wilson loop calculations in the Feynman gauge \cite{Drummond:2007aua,Brandhuber:2007yx}. The latter were shown to reproduce the $n$-gluon one-loop amplitude.\footnote{The fact that the integrals in the one-loop Wilson loop can be rewritten in a form equivalent to the two-mass easy box integrals, after a suitable identification of the regularization parameters, was already pointed out in \cite{Gorsky:2009nv}. } \footnote{We remark that the effective gauge change, from Feynman to Landau, observed above is not surprising. Indeed, as discussed in \cite{Eden:2000mv}, inserting the {gauge-invariant} Lagrangian into a gluon propagator in any gauge brings this propagator to the transverse Landau gauge. So, what happens, at least at one loop, is that if we compute the correlator in dimensional regularization by inserting the Lagrangian, we obtain a result identical with the Wilson loop calculation done in the Landau gauge. This is in contrast with the situation where one calculates the one-loop correlator in the Feynman gauge, but {without insertions}. There the correlator is expressed in terms of more complicated integrals (see \cite{Eden:1998hh}), which coincide with the one-loop scalar box in $D=4$ (off the light cone), but require a special investigation in $D=4-2\ep$ (on the light cone).}

Finally, let us discuss the case of  the one-loop correlator
\begin{equation}\label{kkoo}
\langle {\cal K}(1) {\cal K}(2) O(3) \tilde O(4) \rangle\,,
\end{equation}
involving  two copies of the  Konishi operator \p{Ko}, in addition to two other protected ones.


\vskip 0.2 cm

\begin{figure}[h!]
\psfrag{0}[cc][cc]{}
\psfrag{1}[cc][cc]{$1$}
\psfrag{2}[cc][cc]{$2$}
\psfrag{3}[cc][cc]{$3$}
\psfrag{4}[cc][cc]{$4$}
\centerline{ \includegraphics[width=\textwidth]{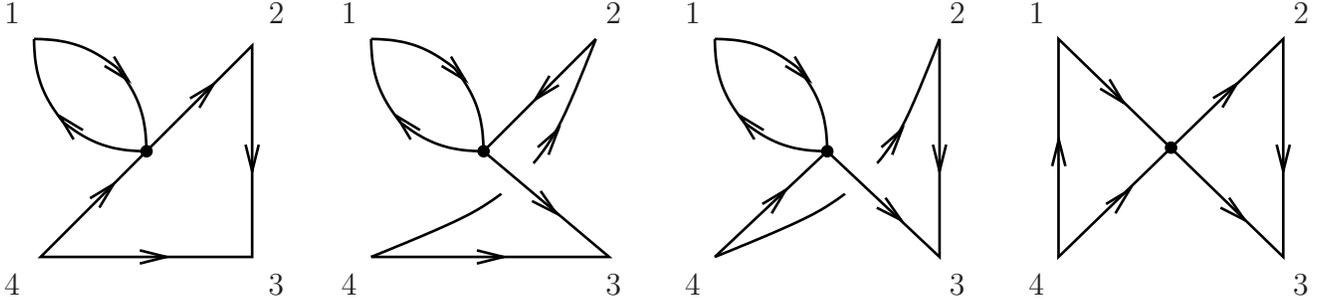} }
 \caption{\small Additional graphs
 with a $\phi^4-$coupling contributing to the correlation function \re{kkoo}. The
 contribution of the first three diagrams is given by the first line in \re{Konishi} while
 the contribution of the forth diagram is given by the last line in \re{Konishi}.} \label{phi4}
 \end{figure}

\noindent The graphs we have listed before
all occur in the same way in this calculation, and in addition we find a
contribution from the insertion of the scalar $\phi^4-$potential. Divided by the tree,
the new terms in the connected part are, up to an overall factor\footnote{The one-loop scalar integral $g(1,2,3,4)$ is taken from \p{defg}, with a $(4-2\ep)-$dimensional measure.}
\begin{eqnarray}\label{Konishi}
&& x^2_{12} x^2_{14} \left( g(1,1,2,4) + \frac{x^2_{23}}{x^2_{24}} g(1,1,2,3) +
\frac{x^2_{34}}{x^2_{24}} g(1,1,3,4) \right)  \\
&+& x^2_{12} x^2_{23} \left( g(1,2,2,3) + \frac{x^2_{14}}{x^2_{13}} g(1,2,2,4) +
\frac{x^2_{34}}{x^2_{13}} g(2,2,3,4) \right)  \nonumber \\
&+& 2 \, x^2_{14} x^2_{23} \, g(1,2,3,4) \rightarrow 0\,, \qquad \mbox{when} \quad
x^2_{i,i+1} \rightarrow 0 \, . \nonumber
\end{eqnarray}

In Figure \ref{phi4} we have depicted the graphs from the first and the last line.
It is important here to work with the bare operators, and to use the stronger
form of the light-cone limit, in which all $x^2_{i,i+1} = \delta$ go to zero at
the same rate. We conclude that the correlator \p{kkoo} involving  unprotected operators also becomes a Wilson loop in the light-cone limit.



\section{The Wilson loop from the correlation function in four dimensions}\label{Wlfcf}

In this section we examine the behavior of the four-dimensional correlation
function \p{nO} as we take the combined light-like limit $x_{i,i+1}^2\to 0$.    We start
from the full renormalized four dimensional correlation function defined at space-like separated points and then
we take these points to approach the vertices of a polygonal null Wilson loop, making
all distances of the form $x^2_{i,i+1}$ light-like. In the language of the previous section, we first remove
the regulator and then we take $x^2_{i,i+1}$ to zero.
We will consider first a weakly coupled theory and operators of dimension two which are constructed from
two fields in the free theory.

As we take the light-like limit the correlator develops some
singularities. Thus we   need to understand precisely the singular
behavior as we approach the light-like limit. If only two points
were becoming light-like separated this limit would be a light-cone
operator product expansion, which is a well explored subject, see e.g. the review
\cite{BKM}. Of course, the basic physics is simple, we have a fast
moving particle going between the different vertices of the polygon.
This fast moving particle is colored and gives rise to a Wilson loop.
In section \ref{cflc}  this particle was moving so fast that its
momentum was above the UV regularization scale, so it behaved like
an ordinary free particle. In the present case it behaves as a more complex
object. In fact, it behaves as a particle that has a color flux tube
attached to it. Since the particle is in the adjoint,   this  leads
in the planar approximation to  two copies of the Wilson loop (one in
the fundamental and one in the anti-fundamental).

The approach to the limit is determined by the symmetries of the theory. For this reason it is convenient to
 focuss  clearly on the symmetries.
The symmetries are those of a Wilson loop with a cusp. There are two relevant non-compact symmetries:  the boost
centered at the cusp and the dilatation also centered at the cusp. These two symmetries commute.
They can be made manifest by choosing appropriate coordinates and an appropriate conformal frame, see \cite{AMtwo}.
We write the original metric as $ds^2 = dx^+ dx^- + dr^2 + r^2 d\varphi^2$. We then multiply it by an
overall Weyl factor $1/r^2$ to get the metric of $AdS_3 \times S^1$. Then in this $AdS_3$ subspace we choose
coordinates
\beqa
ds^2_{R^{1,3}} &=& \Omega^2 (  ds^2_{AdS_3}  + d \varphi^2 ) ~,~~~~~~~~~~~\Omega = r \,, \label{flAdSthree}
\\
ds^2_{AdS_3}  &=& d \tau^2 + d\sigma^2 + 2 d\tau d\sigma \sin 2 \beta  - d \beta^2
\,. \label{AdSthree}
\eeqa

Let us consider first a configuration that can give rise to the square Wilson loop. We choose four points
sitting at $\tau = \pm \tau_0 $ and $\sigma = \pm \sigma_0$ and $\varphi =0$, $\beta =0$. These are four
space-like separated points. As $\tau_0 , \sigma_0 \to \infty$ the  points approach the vertices of a square
light-like polygon. Naively, from the metric in \nref{AdSthree} one finds that the points are moving
infinitely far away. However, due to the conformal factor $\Omega$
 that we have introduced in going from $R^{1,3}$ to the metric in \nref{AdSthree} the points are actually
 getting closer in the $R^{1,3}$ metric.
  Thus, the  coordinates in \nref{AdSthree}
  resolve the region that was giving rise to the divergence. In fact,  they have
transformed the UV singularity into an IR singularity. Of course,  this is a common occurrence in conformal
field theories.
It is useful to compute the propagator for a field of dimension one in these
coordinates. This can be done by writing the general propagator as $\langle \phi(1) \phi(2) \rangle =
{ G(1) G(2) /( Z_1 \cdot  Z_2) }$ where $Z$ are the coordinates of the projective
light-cone in $R^{2,4}$, which obey $Z^2 =0$ and $Z\sim \lambda Z$. $G $ is a function of the $Z$'s of homogeneous
degree one. Then the above correlator is the one that one should use if we ``gauge fix'' the symmetry of
rescaling of $Z$'s by setting $G=1$.  For example, the
ordinary $R^{1,3}$ coordinates are obtained by taking $G= Z^{-1} + Z^4 $ and
 $x^\mu = Z^{\mu}/G$, with $\mu =0,1,2,3$.
The coordinates in \nref{AdSthree} are obtained by taking $\tilde G = \sqrt{ (Z^3)^2 + (Z^4)^2 }$.
Then the coordinates in $AdS_3$ correspond to those on the hyperboloid $ Y^2 =-1$,
$Y^\alpha = {Z^\alpha / \tilde G }$, (with  $\alpha =-1,0,1,2$),  see appendix \ref{coordchanges}
 for more details.
The propagator in the new coordinates takes the form
\be \label{corrnew}
  \Omega(1) \Omega(2) \langle \phi(1) \phi(2) \rangle_{R^{1,3}} =
 \langle \phi(1) \phi(2) \rangle_{AdS_3 \times S^1}  =
 { 1 \over 2 \cosh \tau_{12} \cosh \sigma_{12} -2 } ~,
\ee
where we evaluated it at   $\beta_i = \varphi_i =0$. At these points one also finds that
\be
\Omega = { \tilde G \over G } = r = { 1 \over \cosh \tau \cosh \sigma } ~, ~~~~~~~~ \beta = \varphi =0.
\ee
Since the $R^{1,3}$ correlator is inversely proportional to the square of the distance we can also
use  \nref{corrnew} to convert distances into expressions in the new coordinates.

Any four points that are space-like separated can be mapped via a
conformal transformation to four points which are at the vertices of
a rectangle in the $\tau, \sigma$ plane at $\beta =\varphi=0$ (see
Figure~\ref{squaretausigma}). We can use any coordinate systems to
compute the cross ratios of these points. We find that the cross
ratios are \beqa \label{CRtau}
 |z|^2 &=& { x_{AD}^2 x_{BC}^2 \over x_{AC}^2 x^2_{DB}   } = \left( { 1 - \cosh \Delta \tau  \over
 1 - \cosh \Delta \sigma \cosh \Delta \tau } \right)^2 \propto e^{ - 2 \Delta \sigma }\,,
\\ \notag
|1-z|^2 &=& { x_{AB}^2 x_{DC}^2 \over x_{AC}^2 x^2_{DB}  } = \left( { 1 - \cosh \Delta \sigma \over
 1 - \cosh \Delta \sigma \cosh \Delta \tau } \right)^2 \propto e^{ - 2 \Delta \tau }\,,
\eeqa
where   $\Delta \tau$ and $\Delta \sigma $ are the sizes of the rectangle. We assumed that
both are large $\Delta \tau,\Delta \sigma \gg 1$, but their ratio $\Delta \tau/\Delta \sigma$
is arbitrary.
The corrections to the above approximation are in powers of the cross ratios that are becoming small
and can be neglected for what we will do below.

\begin{figure}[t]
\begin{center}
\includegraphics[width=120mm]{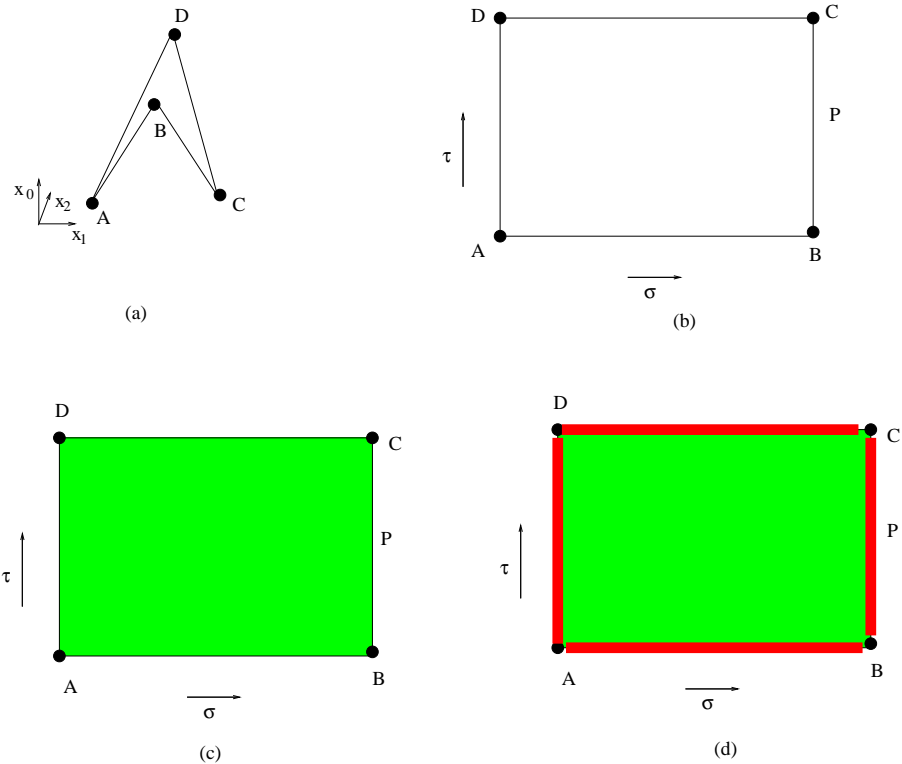}
\end{center}
\caption{ In (a) we see a polygon with nearly light-like lines. The
dots indicate the insertion points of the operators and the lines
represent the propagation of a field. In (b) we see the same in the
coordinates introduced in \nref{AdSthree}. The points form a
rectangle and the light-like limit corresponds to making the
rectangle  bigger and bigger. The lines represent the propagation of
a fast moving particle in the tree level theory. (c) The particles
source a color electric field which is extended on the square. (d)
This color electric field modifies the propagation of the particles
in the full theory. The red lines represent this modification which
happens locally at each line. } \la{squaretausigma}
\end{figure}

We see that as we take $\tau_{12} \to \infty$ in \nref{corrnew} {the propagator scales as} $e^{ - |\tau_{12} | }/\cosh \sigma_{12}$.
The fact that it factorizes into a function of
 $\tau$ and a function  of $\sigma$ implies that the energies are independent of the momentum. Here we
 are defining an energy generator by $\partial_\tau$ and a momentum by $ -i \partial_\sigma$.\footnote{
 It turns out that this is the same as saying that the
twist of a field is independent of the spin, where the twist is defined as $\Delta - S$.}. These
two operators correspond in the conformal algebra to the operators $D\pm M_{+-}$ so that for
the particle moving in the `$+$' direction one of them measures the twist and the other the conformal
spin. For the particle moving in the `$-$' direction it is the other way around.
This is a result in the free theory. Thus the leading
result in the free theory comes from the free propagation of the particles indicated by lines in Figure~\ref{squaretausigma}(b). This is modified as we go to the interacting theory.
Extra singular terms arise precisely due to this modification. Of course, thanks to the factors of $\Omega$ in
translating back to the original coordinates, we get the usual $1/x_{i,i+1}^2$ singularities in the limit.
The  factors of $\Omega$ are determined by the dimensions of the external operators. For the half-BPS operators of dimension two \p{bil}, these factors are fixed once and for all and are used
to produced the tree level answer.

What are the modifications in the interacting theory?
The first and most important modification is {due to the fact that the propagating particles carry color charge in the adjoint representation. They create a color flux  in the adjoint}
that goes between the four lines in Figure~\ref{squaretausigma}(c). This color electric flux has a constant
energy density in the $(\tau, \sigma)-$plane. This is not obvious from
what we said so far. It becomes   more clear if we choose a
Lorentzian picture and continue $\tau \to i t, ~~ \beta \to i \hat
\beta$. In this case we have the two time-like lines that source the
electric field. We consider the situation where these two lines are
at a very big distance in the $\sigma-$direction. The final result of
the analysis in \cite{AMtwo} is that for large $t$ and $\sigma$ one
should think of the dynamics as happening in the two dimensions
spanned by $t,\sigma$ and the flux has constant energy density. The
reason is the following.   The $\varphi$ direction is a circle and
fields can be KK reduced and at long distances only the constant
modes would be relevant. The $ \hat \beta$ direction is non-compact,
but there is effectively a gravitational potential   confines the
electric flux along this direction.%
 \footnote{The energy conjugate to
  $t$ can be viewed as the twist $\Delta -S$ when we analyze high spin operators or operators defined
  along a light-like direction.} The effects of the color electric field
     produce a factor which is proportional to the area of the rectangle in
  Figure~\ref{squaretausigma} (c). The proportionality constant is simply the cusp
   anomalous dimension $\Gamma_{\rm cusp}^{\rm adj}$,\footnote{We are
    normalizing $\Gamma^{\rm adj}_{\rm cusp}$ to be the energy density
   in the $\tau, \sigma$ plane for a flux in the adjoint.   One might be worried that the flux in the
   adjoint would be screened. In the planar approximation it is not. Also in perturbation theory around
   weak coupling it is not screened, even if we are not in the planar approximation.  }
  see \cite{AMtwo}. In the planar approximation $\Gamma_{\rm cusp}^{\rm adj} = 2 \Gamma_{\rm cusp}$, where
  $\Gamma_{\rm cusp}$ is the energy density of a flux in the fundamental.
   Thus we get a contribution of the form $\Gamma^{\rm adj}_{\rm cusp} \Delta \tau \Delta \sigma$.
   Using \nref{CRtau} this can be reexpressed as
  \be
   \Gamma^{\rm adj}_{\rm cusp} \Delta \tau \Delta \sigma
    \sim {\Gamma^{\rm adj}_{\rm cusp} \over 4 } \log |z|^2 \log |1 -z|^2 \sim  { \Gamma^{\rm adj}_{\rm cusp} \over 4 }  \log z \log (1-\bar z )\,.
    \ee
    Note that in the light-cone limit  we have $z \to 0$ and $\bar z \to 1 $\footnote{Note that $z$ and $\bar z$ are complex
    conjugates in the Euclidean theory but they can be both real in the lorentzian theory.}.
  In general, when two consecutive lines are becoming light-like we can see that there will be a factor of
  the form $\log x_{12}^2 \log x_{23}^2 $ from this reasoning.
  So far, we are getting a divergent factor of the form
  \be \label{sqenergy}
  \langle O(1) \cdots O(n) \rangle \longrightarrow
   e^{ - { \Gamma^{\rm adj}_{\rm cusp} \over 4 }  \sum_{i=1}^n \log { x_{i-1,i}^2 \over x_{i-1,i+1}^2 }
   \log {  x^2_{i,i+1} \over x_{i-1,i+1}^2 } }\,.
  \ee
  This is the most important factor, the rest of the terms involve essentially single logs.

\begin{figure}[t]
\begin{center}
\includegraphics[width=120mm]{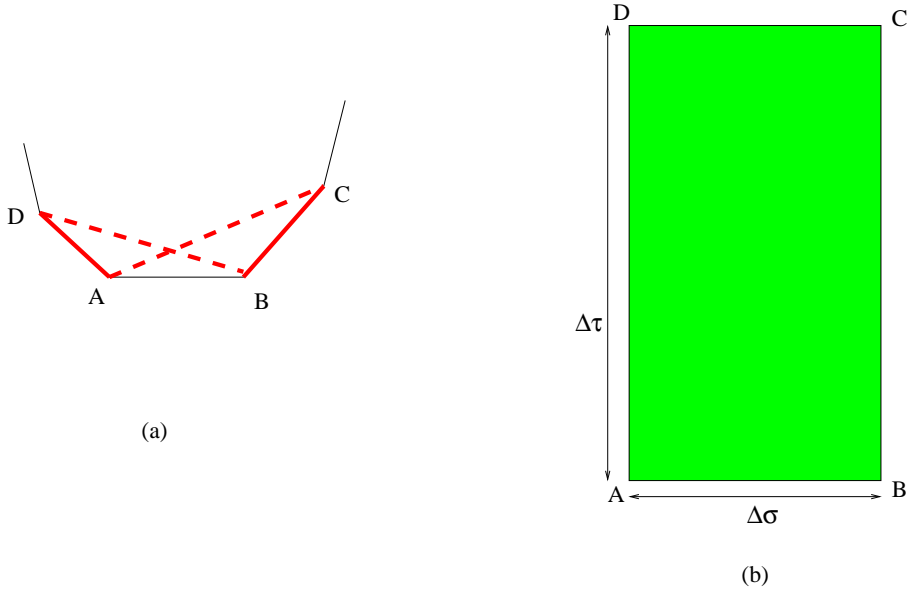}
\end{center}
\caption{ If we concentrate on points $x_i$ and $x_{i+1}$ (here denoted by $A$ and $B$) then we see that their distance in the
$\tau,\sigma$ plane is given by $\Delta \sigma$ and this is given by the cross ratio $u_{i,i+1}$
represented graphically
in (a). Red distances appear in the cross ratio. Solid lines in the numerator and doted lines in the
denominator.
} \la{deltatau}
\end{figure}

  Let us now turn to the second effect that is related to the
  modification of the propagation of the particles in Figure~\ref{squaretausigma} (d) 
  due to their interaction with the color flux.
  There are two {sources of corrections}.
  The first is that the energy of the particles can be modified. Instead of being precisely 1, it can be
  slightly bigger or smaller. If that were the only effect, it would be very easy to take it into account.
   One would need to change the propagator $e^{-\Delta \tau_{i,i+1}}$ to $e^{ - (1 + \tilde g )
   \Delta \tau_{i,i+1} }$. Using the formulas in \nref{CRtau} applied to points $i-1,i,i+1,i+2$ we find that
    \be \label{CRu}
\Delta \tau_{i,i+1} \sim
 - { 1 \over 2 }  \log { x_{i+1,i+2}^2 x^2_{i-1,i} \over x_{i,i+2}^2 x_{i-1,i+1}^2 }  \equiv - { 1 \over 2 } \log u_{i,i+1}\,.
\ee
 We defined the cross ratio $u_{i,i+1}$ which  measures the displacement in the
 $(\tau,\sigma)-$plane between two consecutive points, see Figure~\ref{deltatau}.
 This leads to an extra factor of the form
 \be \label{enshift}
   e^{ \sum_{i=1}^n { \tilde g  \over 2 } \log u_{i,i+1} }\,,
 \ee
 where we used that all the particles that are exchanged have identical energy shifts.
 If we were exchanging uncolored, gauge invariant
  states, this would be the only effect.
  However, in our case, the particles also have an electric flux attached to them. Thus, they behave
  as particles subject to a constant force, and   we expect them to ``bend'' in the presence of this force.
  The details of this bending depend on the modifications to the dispersion relation, etc. However, we
  do not need to work them out explicitly for our purposes.
  For example, let us consider the particle labeled by P in Figure~\ref{squaretausigma}. It
  moves along the $\tau$ direction. It would not move along the $\sigma-$direction if there were no
  electric field. In the free theory the particle has a discrete spectrum of energies (or twists), $ E=1,2$.
  These values of the twist correspond to different excitations in the theory. For example, a scalar or some
  components of the gauge fields  have twist one, etc.
  Let us concentrate on the particle of energy $E=1$, which is the dominant one in our limit.
  It is important
  to note that even though the energies are discrete, there is an infinite degeneracy at each energy level.
  One can label this degeneracy in different ways. For example, we can label the particle by its momentum
  along $\sigma$. The energy is independent from the momentum in the free theory. As we add the electric
  field we break the degeneracy. It is useful to understand how this
  degeneracy is broken. Let us write the energy as $E =1 + \epsilon$. With enough
  patience we could find the energy eigenfunctions $\psi_{\epsilon}(\sigma)$.
  The important observation is that the original translation symmetry of the
  problem plus the fact that the electric field is a constant we can deduce that the wavefunctions
  for different energies are all related to each other
  \be
  \psi_{\epsilon}(\sigma) = \psi_0\left( \sigma -  { \epsilon \over   \Gamma^{\rm adj}_{\rm cusp}}  \right)
  \ee
  This is a fact that might be familiar to readers who have found the energy eigenstates for a particle in
  a constant gravitational field. A shift in $\sigma$ translates, through the electric field, into a shift
  in the energy. We can wonder about the choice of origin in $\sigma$. The electric field is to the left
  of the particle $P$, see Figure~\ref{wavefunctions}. Of course this shift of origin will
  just lead to constant shift in the energy, or a shift in $\tilde g$ in \nref{enshift}.
  Also, when we took into account the leading effects of
  the color electric field, we have cut off  its  contribution at the position of the insertions of
  the operator. Thus, we should use this value of $\sigma$ as an origin, and also to set the zero of the energy.

  Small constant shifts lead to the terms linear in $\tau$ already taken into account in \nref{enshift}.
   However, the fact that the operator produces a state with a range of
   energies  produces a new effect which we now describe.

\begin{figure}[t]
\begin{center}
\includegraphics[width=45mm]{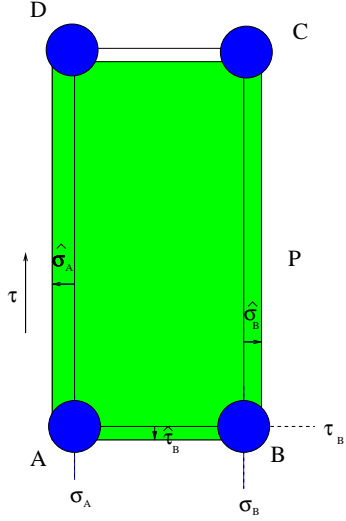}
\end{center}
\caption{  Each operator creates states with a range of energies.  A state with a given energy is
represented a a straight line which extends up to a new location. For example, the operator $B$ is
inserted at $\sigma_B$, but the vertical line corresponding to particle $P$ extends to
$\sigma = \sigma_B + \hat \sigma_B$. This extent in the $\sigma $ direction is just a nemonic device
that tells us what the energy of the state is.  There are similar particle propagating from $A$ to $B$
characterized by $\hat \tau$, etc.
} \la{wavefunctions}
\end{figure}

  Let us consider the particle $P$ that propagates between points $B$ and $C$ in Figure~\ref{wavefunctions}.
    We denote by $\hat \sigma_B $ the deviation of $\sigma$ from the position of the insertion of
    the operator which we take at $\sigma= \sigma_B$,
     $\hat \sigma_B = \sigma -\sigma_B $. Instead of labeling the intermediate states by their energies
     we use $\hat \sigma_B$ as a label. The advantage is that we can think of the total energy of the
     configuration as $\Gamma^{\rm adj}_{\rm cusp} \hat \sigma_B$, where again we have subtracted the part of
     the flux energy that we have already taken into account. Thus we can label the wavefunction as
     $\psi_{\hat \sigma} (\hat \sigma') = \psi_0(\hat\sigma' - \hat \sigma )$. We will be evaluating
     this wavefunction at $\hat\sigma' =0$ which is the insertion point of the operator. Thus we
     simply get $\psi_0(-\hat \sigma)$. The operator creates a pair of particles with a range of energies.
     For the operator at $B$, one of them propagates in the $\tau$ direction, towards $A$ and the other
     in the $\sigma$ direction towards $C$. We denote by
   $C_B(\hat \sigma_B,\hat \tau_B)$ the amplitude that the operator creates a pair of particles with the
   corresponding energies, labeled here by $\hat \sigma_B$ and $\hat \tau_B$,
     see Figure~\ref{wavefunctions}.
   There is a similar factor for  the operator sitting at point $C$.
   Thus we will get a contribution which goes as
   \be
    \int d \hat \sigma\, C_C( \hat \sigma  , \hat \tau_C) C_B( \hat \sigma , \hat \tau_B )\,
    e^{ -   \Gamma^{\rm adj}_{\rm cusp} |\tau_{BC}| \hat \sigma } \psi_0(-\hat \sigma) \psi_0^*(-\hat \sigma)\,.
    \ee
    This formula can be interpreted as the expansion of the
    transition amplitude for a scalar particle to go from point B to C in terms of energy (and conformal spin)
    eigenstates.
     Here we have set $\hat \sigma_B = \hat \sigma_C = \hat \sigma $ since these are energy eigenstates.
      Note that
     the only term that is large here is $\tau_{BC}$ which is diverging in the light-cone limit.
     There are similar terms involving the propagation of the other particles.

     For the case of the square, the final contribution from all such factors has the form
     \beqa \label{Jdef}
    J &=&   \int d \hat  \sigma d \hat\sigma' d\hat \tau d\hat \tau'
      C_A( \hat \sigma , \hat \tau) C_D(\hat \sigma , \hat \tau ') C_B(\hat \sigma',\hat \tau' )
       C_C(\hat \sigma' , \hat \tau' )
    \\ \notag
   & \times &  e^{ - \Gamma^{\rm adj}_{\rm cusp}  |\tau_{AD}| \hat \sigma }  e^{ - \Gamma^{\rm adj}_{\rm cusp}  |\sigma_{AB}| \hat\tau}
    e^{ - \Gamma^{\rm adj}_{\rm cusp}  |\tau_{BC}| \hat \sigma'}  e^{ - \Gamma^{\rm adj}_{\rm cusp} | \sigma_{DC}| \hat \tau' }
    \\ \notag
    &\times &
    \psi_0(-\sigma) \psi_0^*(-\sigma)\psi_0(-\sigma') \psi_0^*(-\sigma')
   \psi_0(-\tau) \psi_0^*(-\tau) \psi_0(-\tau') \psi_0^*(-\tau')\,.
   \eeqa
In deriving this relation we have only assumed very general symmetry
properties of the theory. Thus, we expect that it should
continue to be true for all values of the coupling.  We did assume
that we created a state that is characterized by a single label, the
energy, or equivalently $\hat \sigma$. This is definitely true for
the creation of  twist one excitations in the perturbative theory.
This is what is created by an operator \p{2f} that has two fields. The
coefficients $ C(\hat \sigma, \hat \tau) $ depend on the details of
the theory and we do not have any prescription for computing them in
general. We can replace in \nref{Jdef} the distances in the $(\tau,\sigma)-$plane
by $u_{i,i+1}$ using \nref{CRu}.

The extra factor $J$ looks complicated. However, it only starts contributing at two loops.
We will see that for two loops we can actually compute explicitly
the functions appearing in it by relating them to the functions that
appear in the ordinary operator product expansion. We will do the
 explicit comparison in section \ref{twoloopcheck}.

{To summarize our discussion, the correlation function in the
light-cone limit that we are considering takes the following form}
\beqa \label{OPEWilson}
{G_n  \over  G_n^{\rm tree} }
&  = &
   e^{ - {\Gamma^{\rm adj}_{\rm cusp} \over 4 }  \sum_{i=1}^n \log { x_{i-1,i}^2 \over x_{i-1,i+1}^2 }
   \log {  x^2_{i,i+1} \over x_{i-1,i+1}^2 }  + {\tilde g (\lambda) \over 2 }
\sum_{i=1}^n  \log u_{i,i+1}  }  \, J \,  W_{\rm ren}^{\rm adj}\,,
\eeqa
with
\beqa \label{Jgendef}
J &=& \int \prod_{i=1}^n d\hat \sigma_i \,C_i(\hat \sigma_i , \hat \sigma_{i+1} ) |\psi_0(-\hat \sigma_i) |^2
 e^{ { 1 \over 2 } \Gamma^{\rm adj}_{\rm cusp} \sum_{i=1}^n \hat \sigma_i \log u_{i,i+1} }\,,
\eeqa
where we denoted $\hat \tau_i$  by $\hat \sigma_i$ to simplify the notation.
 Recall that $u_{i,i+1}$ vanish in the same light-cone limit.
Here $W^{\rm adj}_{\rm ren}$ is the renormalized
 Wilson loop in the adjoint.
It is  renormalized by cutting off the divergencies in a
way that depends on the distances $x_{i,i+1}^2$ not being quite light-like and subtracting the terms
that appear multiplying in \nref{OPEWilson}.  {Here we are assuming that we have BPS operators of dimension
two that are not renormalized.}
 This is a very particular way of regularizing the Wilson loop.
In order to compare to other ways of regularizing the Wilson loop we need to understand how to translate to
a different regularization.
We will do this translation though a rather indirect route that works nicely when the number of points, $n$,
is not a multiple of four. We leave the case where $n$ is a multiple of four to the future.

Let us start from the {light-like} Wilson loop in dimensional regularization, for example.
Then we can write the result as
\beqa \la{resdim}
W^{\rm adj}  &=& e^{\rm Div} \,  W^{\rm adj}_{\rm finite} ~,~~~~~~~ W^{\rm adj}_{\rm finite} =
e^{\Gamma^{\rm adj}_{\rm cusp} A_{\rm BDS-like} } W_{\rm conformal}\,,
\eeqa
{where the divergent part is given by}
\beqa \la{divpart}
{\rm Div} &=&  - \sum_{i=1}^n { \Gamma^{\rm adj}_{\rm cusp} \over 8 } \left( \log ({  \mu^2   x_{i,i+2}^2 ) } \right)^2 +
 g  \log ({ \mu^2
x_{i,i+2}^2  }) + [{\rm poles}]\,,
\eeqa
and [poles] denotes terms involving simple and double poles in $\epsilon$.
The finite piece $W^{\rm adj}_{\rm finite} $ obeys the Ward identities for anomalous special conformal symmetries \cite{Drummond:2007au}.
If $n$ is not a multiple of four, we can construct a unique solution of the conformal Ward identities that
involves only next to nearest neighbor  distances,  $x_{i,i+2}^2$. This
unique expression is what we called $A_{\rm BDS-like}$ in \nref{resdim}.
 Its explicit form can be found in formula (A.7) of \cite{AGM}%
\footnote{Actually, they differ by a factor of two and a sign $A^{\rm here}_{\rm BDS-like} = -2 A^{\rm in ~\cite{AGM}}_{\rm BDS-like}
  $.}.
Then the rest, which we called here $W_{\rm conformal}$, is a conformal invariant function of $x$'s.

Now, we can do the same trick in the above discussion. One aspect
that was not particularly nice in the above discussion was the fact
that we had explicit distances appearing in the formulas.  In fact,
if we factor out the tree expression \nref{OPEWilson}, the
correlator becomes conformal invariant. Thus, it seems ugly to break
the conformal invariance in order to extract the Wilson loop. For
this reason, it is convenient to preserve conformal invariance
throughout the computation. This can be achieved as follows. First
let us understand in more detail what happens as one of the
distances $x_{i,i+1}^2$ goes to zero while holding the rest small
but fixed. In that case we can map the four points $i-1,i,i+1,i+2$
to the vertices of a rectangle in $\tau, ~\sigma$ plane, see Figure~\ref{deltatau}.  Then the limit $x^2_{i,i+1} \to 0$ corresponds to
$\Delta \tau \to \infty$ and the area grows as
\be \label{singledis}
\log G_n \to \Gamma^{\rm adj}_{\rm cusp} \Delta \tau \Delta \sigma = {
\Gamma^{\rm adj}_{\rm cusp} \over 4 }  \log x^2_{i,i+1} \log
u_{i,i+1}\,,
\ee
where we used \nref{CRu}. This is valid up to terms
that remain finite when $x^2_{i,i+1} $ goes to zero. Of course, the
expression \nref{sqenergy} obeys \nref{singledis}. We now would like
to write an expression which involves only cross ratios which also
obeys \nref{singledis}. This is easy to do when $n $ is not a
multiple of four. In that case,
  a given distance   $x_{i,i+1}^2 \to 0$  can be promoted to
  a unique cross-ratio (if $n \not = 4 k$) that involves the distance $x^2_{i,i+1}$ and
next-to-neighboring distances of the form $x^2_{j,j+2}$. We denote by $\chi_{i,i+1}$ this unique cross-ratio.
Then one also finds that $u_{i,i+1} = \chi_{i-1,i}\chi_{i+1,i+2}$.
Thus each cusp gives rise to a term $ - { \Gamma_{\rm cusp}^{\rm adj} \over 4 } \log \chi_{i-1,i} \log \chi_{i,i+1} $.
Then we can rewrite the limit as\footnote{Here we assumed that the operators are BPS so that dividing by
the tree correlator we get a conformal invariant answer.}
\beqa
{ \langle O(1) \cdots O(n) \rangle \over
\langle O(1) \cdots O(n) \rangle_{\rm tree} }
  &=& {\cal F}(\mbox{cross-ratios})   
  \nn\\
  &= & \la{finexpan}
e^{ -{  \Gamma^{\rm adj}_{\rm cusp} \over 4 } \sum_{i=1}^n\log \chi_{i-1,i} \log \chi_{i,i+1} + {\tilde  g(\lambda)
\over 2 }
\sum_{i=1}^n  \log u_{i,i+1} }  \, J\,  W^{\rm adj}_{\rm conformal}\,.
 \eeqa
In writing \nref{finexpan}
we have asserted that $W_{\rm conformal}$ is indeed the same as $W_{\rm conformal}$ in \nref{resdim}.
More precisely, in a planar theory we would get $W^{\rm adj}_{\rm conformal} = (W_{\rm conformal})^2$, where
$W_{\rm conformal}$ is the Wilson loop in the fundamental.
 The reason is that
both are a  renormalized version of the Wilson loop which has been made conformal invariant
by the use of next-to-nearest neighboring distances. Thus we expect that both should coincide.
In the special case of the four point function, $n=4$,
the expression \nref{finexpan} continues to be valid after replacing $\chi_{12} \chi_{34} =
u_{2,3} = |z|^2 $, $\chi_{23} \chi_{41} = u_{34} = |1-z|^2 $. The cases where $n$ is other multiples of
four is left for the future.

In fact, we will check below that the two expressions for $W_{\rm conformal}$ coincide at one loop.
At one loop the factor $J$ can be set to one. The reason is that any single logarithm appearing in $J$
can be absorbed in a redefinition of the function $\tilde g$. Only at two loops we expect to see the appearance
of the factor $J$. In fact, we consider the four point correlation function (which is the only available
full correlator  at two loops) and we check the appearance of terms that can be interpreted as arising from
the factor $J$. We will perform these two checks in the next two subsections.

 The attentive reader might be surprised by a certain factor of  two for the leading double logarithmic terms
  in the expression for the correlation function $G_n$, Eq.~\re{sqenergy},
   and light-like Wilson loop in the adjoint representation $W^{\rm adj} $, Eq.~\nref{divpart}.
Namely, if we set the distances $x^2_{i,i+1} = \mu^{-2}$ then  the coefficient
in front of double logs in the two expressions  differ
by a factor of two.  The origin of this factor of two is explained in appendix
 \ref{FactorTwo}. (This two is unrelated to the one that arises when we go from
  the adjoint to two copies of the fundamental in the planar limit.)

\subsection{ One-loop checks}

In this subsection we check {by an explicit one-loop calculation} that the light-cone limit of the $n-$point correlators \p{defco} has the form \p{finexpan}, and evaluate the
one-loop expression for $W_{\rm conformal}$. As we mentioned above, the factor $J$ can be set to $1$ at one loop.

The one-loop correlators of BPS operators   were computed in \cite{GonzalezRey:1998tk,Eden:1998hh,Drukker:2009sf}.
In our limit we get
\begin{align}\label{1lo}
  \lim_{x^2_{i,i+1}\to 0}
\left( G_n/G^{\rm tree}_n \right)   &=   - \frac{i
a}{4   \pi^2} \lim_{x^2_{i,i+1}\to 0} \sum_{k,l} \int
d^4 x_0  \frac{ x_{k,l+1}^2 x_{k+1,l}^2- x_{kl}^2
x_{k+1,l+1}^2}{x_{k,0}^2x_{k+1,0}^2 x_{l,0}^2x_{l+1,0}^2} \nonumber \\
& =  -   \frac{a}{2} \lim_{x^2_{i,i+1}\to 0} \sum_{k,l}
\left(x_{k,l+1}^2 x_{k+1,l}^2- x_{kl}^2 x_{k+1,l+1}^2 \right)
g(k,k+1,l,l+1)\,  ,
\end{align}
where $a= {g^2 N_c}/({8\pi^2})$ is the 't Hooft coupling constant, and the integral $g(1,2,3,4)$ (the dual space version of the one-loop scalar box) is defined as
\begin{equation}\label{defg}
g(1,2,3,4)   =   \frac{i}{2 \pi^{2}}
\int\frac{d^{4} x_0}{x^2_{10}   x^2_{20}   x^2_{30}   x^2_{40}}\,  .
\end{equation}
As long as the outer points are kept in generic positions, $x^2_{i,i+1} \neq 0$, this integral is finite and conformally covariant in $D=4$ dimensions. This allows us to write it down as a function of two conformal cross-ratios
\begin{equation}\label{defg4}
g(1,2,3,4)   =    \frac{1}{x_{13}^2 x_{24}^2}
\Phi^{(1)}(x,y) \,   , \quad \text{with} \quad x    =
\frac{x^2_{14} x^2_{23}}{x^2_{13} x^2_{24}}\,    , \quad y
  =   \frac{x^2_{12} x^2_{34}}{x^2_{13} x^2_{24}}\,    ,
\end{equation}
where the two-variable function $\Phi^{(1)}$ is given by  \cite{davussladder}
\begin{eqnarray}\label{davphi}
\Phi^{(1)}(x,y) & = & \frac{1}{2} \int_0^1 d\xi \frac{\log(y/x) + 2
\log\xi}{y   \xi^2 + (1 - x - y)   \xi + x}\, .
\end{eqnarray}

{Let us now examine the light-cone limit  $x^2_{i,i+1} \to 0$.} In Eq.~(\ref{1lo}) the arguments of the $g$ integrals are
pairwise adjacent, $(x_k,x_{k+1})$ and $(x_l,x_{l+1})$. This implies that,
{in the light-cone limit  $x^2_{i,i+1} \to 0$,} at least one of the two cross-ratios  
{on which the $g-$integral depends}
goes to
zero in this limit. If in addition $l = k \pm 2$, all
points are adjacent, in which case both cross-ratios tend to zero. The leading singular behavior of $\Phi^{(1)}(x,y)$ when $y \to 0$ is given by
\begin{eqnarray}
\lim_{y \to 0}\Phi^{(1)}(x,y) & = &
 \frac{1}{2} \int_0^1 d\xi \frac{\log(y/x) + 2
\log \xi }{(1 - x)   \xi + x}   +   O(y) \nonumber\\ & = &
-\frac{1}{2(1-x)} \left[\log x  \log y  - 2   \mathrm{Li_2} \left(1 -
\frac{1}{x}\right)   -   \log^2(x) \right]   +   O(y)\,  . \label{limy}
\label{glimy}
\end{eqnarray}
We will need (\ref{limy}) for $n\geq 6$, while for $n=4,5$ the double limit $x,y
\to 0$ is relevant:
\begin{equation}
\lim_{x,y \to 0}\Phi^{(1)}(x,y)   =   - \frac{1}{2} \log x  \log y   -
\frac{\pi^2}{6}   + O(x)   +   O(y) \label{glimxy}\,  .
\end{equation}
In the simplest case of four points we have
\begin{eqnarray}
\lim_{x^2_{i,i+1}\to 0} \left( G_4/G^{\rm tree}_4
\right)  &=& 2   a   x^2_ {13}
x^2_{24}   \lim_{x^2_{i,i+1}\to 0} g(1,2,3,4) \nonumber \\
  &=& -   a   \left[ \log x
\log y   +   \frac{\pi^2}{3} \right] + O(x) + O(y)\,    . \label{explg4}
\end{eqnarray}
If we put
\begin{equation}\label{idreg}
x_{i,i+1}^2   =  1/\epsilon_c^{2}   \rightarrow   0\,,
\end{equation}
where $\epsilon_c$ is the regulator from Appendix~A in \cite{AGM}, we obtain
\begin{eqnarray}\nonumber
\lim_{x^2_{i,i+1}\to 0} \left( G_4/G^{\rm tree}_4
\right) & = & -   a   \left[ 8   A_\mathrm{div} -
\log^2\left(\frac{x^2_{13}} {x^2_{24}}\right)   +
\frac{\pi^2}{3}   \right] \\ & = & -   a   \left[ 8
A_\mathrm{div}   -   2   A_\mathrm{BDS}   +   4\times \frac{5}{2}
  \zeta(2)   \right]\,, 
\end{eqnarray}
where $A_\mathrm{div}$ is defined in equation (A.6) in
\cite{AGM} and at four points $A_\mathrm{BDS}$ is essentially the
$\log^2$ term.\footnote{By definition, $A_\mathrm{BDS}$ is the finite part in the $n-$gluon one-loop MHV amplitude and is given in \cite{Bern:1994zx}.} This term is finite but obviously not conformally invariant.
On the other hand, the correlator $G_4$ in $D   =   4$ is conformal
before putting $x_{i,i+1}^2   =    1/\epsilon_c^{2} $; it is a function of
cross-ratios. But we can do better: by adding and subtracting the
non-conformal expressions $A_\mathrm{BDS}$ from $A_\mathrm{\rm BDS-like}$  (the latter is defined in (A.7) in \cite{AGM})
and putting it together with the other non-conformal piece $A_{\rm div}$, we can produce a manifestly conformal ``remainder'', which is the one-loop value of the factor $W_{\rm conformal}$ from \p{finexpan}. In what follows we
confirm this explicitly for $n=5,6,7$ and claim that it works for all
$n \neq 4 k$.

In the limiting prescription \p{idreg} we have put all small distances
equal. On the other hand, we may ask whether any given small
distance $x_{i,i+1}^2$ can be uniquely promoted to a cross-ratio
using only members of the next-to-nearest neighbor set $\{x_{j,j+2}^2\}$, as is suggested by
the form of $A_\mathrm{div}$ in (A.7) in
\cite{AGM}. In order to answer
this, one simply solves a linear system on the exponents of the
$\{x_{j,j+2}^2\}$ in a general ratio, requiring the conformal
weight to cancel out at each point. No solution exists when the number of points is
$n   =   4 k$. For all other values of $n$ we find
\begin{align}\notag
    n = 4k +1: \qquad & \chi_{i,i+1} = \frac{ x^2_{i,i+1} x^2_{i+2,i+4} \cdots x^2_{i-3,i-1}}{x^2_{i,i+2} x^2_{i+4,i+6}\cdots
x^2_{i-1,i+1} }\\
n = 4k +3:  \qquad & \chi_{i,i+1} = \frac{ x^2_{i,i+1} x^2_{i+3,i+5} \cdots x^2_{i-4,i-2} }{
x^2_{i+1,i+3} x^2_{i+5,i+7}\cdots x^2_{i-2,i} }   \label{chivar} \\
n = 4 k +2:  \qquad & \chi_{i,i+1} = \frac{ x_{i,i+1}^2 }{ \sqrt{ R_i R_{i+1} } }\,, \qquad R_{i} = \frac{ x^2_{i,i+2} x^2_{i+4 , i+6 } \cdots  x^2_{i-2,i} }{x^2_{i+2,i+4 } \cdots x^2_{i-4,i-2} }\,. \notag
\end{align}
We can now check that { at one loop}
\begin{align}\label{chiexp}
  \lim_{x^2_{i,i+1}\to 0} & \left( G_n/G^{\rm tree}_n
\right)    =
-   a   \left[ \sum_{i=1}^n \log \chi_{i,i+1}
  \log \chi_{i+1,i+2}  -   2  ( A_\mathrm{BDS}   -
A_\mathrm{\rm BDS-like} )  +  n \times \frac{5}{2}   \zeta(2)
\right]. 
\end{align}
Since (\ref{1lo}) contains only $g(k,k+1,l,l+1)$, the arguments
of $g$ are always at least pairwise adjacent. This guarantees that the asymptotic expansions
(\ref{glimy}), (\ref{glimxy}) are sufficient to obtain the
one-loop correlators for all $n$. The last equation can easily be
verified up to rather large values of $n \neq 4 k$.


Explicitly, for $n=5,6,7$ we find the following results for the limit \p{chiexp}:
\begin{eqnarray}
\lim_{x^2_{i,i+1}\to 0}\frac{G_5}{G_5^{\rm tree}} &=& - a \left[ \sum_{i=1}^5
\log\chi_{i,i+1}   \log\chi_{i+1,i+2}   +   5
\frac{\pi^2}{3} \right],  \nonumber\\
\lim_{x^2_{i,i+1}\to 0}\frac{G_6}{G_6^{\rm tree}}  &=&   -   a \left[   \sum_{i=1}^6
\log\chi_{i,i+1}   \log\chi_{i+1,i+2}   +   6
\frac{\pi^2}{3} +   \sum_{i=1}^3 \left(\frac{1}{2} \log^2 u_i
+   \mathrm{Li}_2(1 - u_i) \right)\right], \nonumber\\
\label{G7}
\lim_{x^2_{i,i+1}\to 0}\frac{G_7}{G_7^{\rm tree}} & = & -   a \Biggl[   \sum_{i=1}^7
\log\chi_{i,i+1}   \log\chi_{i+1,i+2}   +   7
\frac{\pi^2}{3} +   \sum_{i=1}^7 \left(\frac{1}{2} \log^2 u_i
+   \mathrm{Li}_2(1 - u_i) \right) \nonumber \\ & &  \phantom{- 8
  a \Biggl[  } + \sum_{j>i} \frac{1}{2}   c_{ij} \log u_i
\log u_j    \Biggr]\,.
\end{eqnarray}
In the last relation the coefficient $c_{ij}$ is equal to:
\begin{equation}
c_{ij}   =   0 \; : \; d(i,j)   =   1   , \qquad c_{ij}   =
  -1 \; : \; d(i,j)   =   2   , \qquad c_{ij}   =   1 \; :
\; d(i,j)   =   3 \,   ,
\end{equation}
where $d(i,j)$ is the shortest cyclic distance of two corners of the
heptagon, so $d(2,7) = 2$, etc. The cross-ratios are defined by
\begin{equation}
u_i   =   \frac{x^2_{i,i+4}   x^2_{i+1,i+3}}{x^2_{i,i+3}
x^2_{i+1,i+4}}\,   .
\end{equation}
The finite parts in (\ref{G7}) are the explicit
versions of $ - 2 (  A_\mathrm{BDS} -    A_\mathrm{\rm BDS-like} )$ for these values of $n$.

\subsection{ A two-loop check for the four point correlator}
\label{twoloopcheck}

In this subsection we consider the four-point correlation function at two loops. We { take}
 the planar limit and  set $\Gamma^{\rm adj}_{\rm cusp} = 2 \Gamma_{\rm cusp}$ in all formulas.
The two-loop correction to the four-point correlation function was computed in \cite{Eden:2000mv,Bianchi:2000hn} and it is given
by \footnote{The relation between the variables used in the previous subsection and this one is $x=z
\bar z$, $y=(1-z)(1-\bar z)$.} (with $\lambda=g^2 N_c$)
\begin{align}
\label{ratio}
{ G_4 \over
G_4^{\rm tree} } &=1-{\lambda\over 8\pi^2} \Phi_1(z,\bar z)+{\lambda^2 \over 16 \pi^4}\bigg\{{2+2 z \bar{z}-z-\bar{z} \over 16} [\Phi_1(z,\bar{z})]^2\\
&+  {1 \over 4( z-\bar{z})}\left(\Phi_2(z,\bar{z})-\Phi_2(1-z,1-\bar{z})-\Phi_2({z \over z-1},{\bar z \over \bar z-1}) \right)\bigg\}\,, \nonumber
\end{align}
where the notation was introduced for
\begin{eqnarray}
\Phi_1(z,\bar{z})={1 \over z-\bar{z}}\left(2 {\rm Li}_2(z)-2 {\rm Li}_2(\bar z)+\log(z \bar z)\log {1-z \over 1-\bar z} \right), \\
\Phi_2(z,\bar{z})= 6 \left( {\rm Li}_4(z)-  {\rm Li}_4(\bar z) \right)-3 \log(z \bar z) \left( {\rm Li}_3(z)-  {\rm Li}_3(\bar z) \right)+\cr
+ {1 \over 2} \log^2(z \bar z) \left( {\rm Li}_2(z)-  {\rm Li}_2(\bar z) \right)\,.
\end{eqnarray}
 We can check that the logarithm of the ratio (\ref{ratio}) has the following expansion, up to constant terms, as
 we take $z \to 0^+$ and $\bar z \to 1^-$
\beqa
 && \notag
 \log \left( { G_4 \over
G_4^{\rm tree} }\right)  \sim    -
  { \Gamma_{\rm cusp}  \over 2   } \log z \log (1-\bar z)  -\frac{B}{2} (\log z + \log (1- \bar z ) )   \\ 
 & &
\hspace*{50mm}   + {  ( 2 \Gamma_{\rm cusp})^2 \pi^2 \over 96  } [  (\log z  )^2 + ( \log (1-\bar z) )^2 ]  \,,   \label{secondline}
\eeqa
where
\beqa
 &&  2 \Gamma_{\rm cusp} =
 { \lambda \over 2 \pi^2 } - { \lambda^2 \over 96 \pi^2 } +
 \cdots,~~~~~B= \frac{3}{32 \pi^4}\zeta_3 \lambda^2+\cdots\,,
 \eeqa
 Note that equation \p{secondline} should be trusted only to second order in $\lambda$.
 
 { Let us now compare \p{secondline} with \nref{finexpan}.}
The first line in \p{secondline} corresponds to the terms in \nref{finexpan} involving the cusp anomalous dimension.
 The second line  in \p{secondline} contains double logarithmic
  terms which we would like to interpret as coming from the factor
 $ J$ in \nref{finexpan}.
 For this purpose, we would like to compute the wave-functions $C(\hat\sigma, \hat \tau)$.
 Since we will be looking at terms that involve two logs and two loops, it is clear that we only need
 $C$ at tree level. At tree level the wave function factorizes
  $C = C_T(\hat\sigma ) C_T(\hat \tau)$, since the operator \p{2f} is the product
 of two fields and each field creates one particle.
 Thus, we are interested in terms of the form $u(\hat \sigma ) \equiv [C_T(\hat \sigma )]^2
 \psi_0(-\hat \sigma) \psi_0^*(-\hat \sigma)$.
 More explicitly, the $ J$ factor \p{Jgendef} can be expressed in terms of the Fourier (or Laplace) transform of $u$, which
 we denote by $v(k)$:
 \beqa  \notag 
 J &=& \left[ \int d\hat \sigma u(\hat \sigma ) e^{ \hat \sigma \Gamma_{\rm cusp} \log z  }
   \int d\hat \tau u(\hat \tau ) e^{  \hat \tau \Gamma_{\rm cusp} \log z  } \right]^2
  \\
   \la{Jfour}
     &=& \bigg[ v\big( -i \Gamma_{\rm cusp} \log z \big)v\big(-i \Gamma_{\rm cusp} \log (1-\bar z)\big) \bigg]^2
  \eeqa
with
\begin{align}
v(k) \equiv \int d\hat \sigma e^{i k \sigma} u(\hat \sigma )
\end{align}  
  
The $v-$factors can be computed indirectly by considering a slightly different problem which involves precisely
 the same functions.
 Namely, consider the one-loop four-point correlation function and take the light-cone OPE limit { corresponding to  $z \to 0^+$ with $\bar z$ kept finite.}  
 {We then extract the contribution from each given conformal spin to the OPE and take the limit where $\bar z \to 1$. } In this limit the spin that contributes most to the OPE
 is large. We will then make the same sort of approximations that we discussed above to represent the answer.
 This computation is possible thanks to the expressions in \cite{Dolan:2004iy},
 which spell out the contribution
 of each conformal towers to the OPE.

\begin{figure}[t]
\begin{center}
\includegraphics[width=30mm]{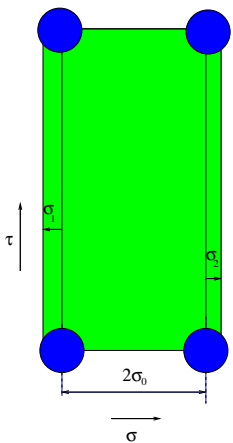}
\end{center}
\caption{ We consider the operator product expansion with $z\to 0$ and $ 1-\bar z$ very small but fixed.
In this regime $\sigma_0$ is large but $\Delta \tau$ is going to infinity. Thus the states that propagate
in the OPE propagate along the $\tau$ direction in this picture. The width of the rectangle represents the
energy of the state. The circles represent the fact that the operator creates states with a range of energies,
so that the energy of the intermediate state depends not only on $\sigma_0$ but also on $\hat \sigma_1$,
$\hat \sigma_2$ which are the extra energies of the propagating particles at the two ends.
} \la{opeapprox}
\end{figure}

We expect that the limit $\bar z \to 1$ should be dominated by the large spin contributions to the OPE.
For reasons that are completely similar to the ones given above,
 we expect that the  large spin  contribution to the OPE for the four-point correlation  function should look
like
\beqa\notag
G_4  &\sim&  e^{ -2  \Gamma_{\rm cusp} \Delta \tau (2 \sigma_0) }
 \int d  \hat \sigma_1 u(\hat \sigma_1) \int d\hat \sigma_2 u(\hat \sigma_2 )
 e^{ - 2  \Gamma_{\rm cusp} \Delta \tau  (\hat \sigma_1 + \hat \sigma_2 ) }
 \\\notag
 & = &
 e^{ -2  \Gamma_{\rm cusp} \Delta \tau (2 \sigma_0) }  \, [ v ( i 2 \Gamma_{\rm cusp} \Delta \tau ) ]^2
\\
 & = & \int d\sigma_s
 e^{ -2  \Gamma_{\rm cusp} \Delta \tau (2 \sigma_0 + \sigma_s ) }  \int d \hat \sigma_1 u(\hat \sigma_1)
   u(\sigma_s - \hat \sigma_1 )\,,
 \la{thirdlineF} \eeqa
where $2 \Delta \tau \sim - \log z $, $4 \sigma_0 \sim - \log (1- \bar z )$ and $v$ is the Fourier transform of $u$,
as in \nref{Jfour}. This formula is derived as follows.
  $\tau$ and $\sigma_0$ characterize the points where the operators are inserted.
The operators create particles at the center of the blue circles in Figure~\ref{opeapprox}. The  particle states
can be expanded in energy eigenstates for the problem of a particle which has a
 color electric adjoint flux ending on it.
The energy eigenstates can be labeled by a position $\hat \sigma_1$. The blue circles represent the overlap
between these wave functions and the field operator. The filled rectangle in Figure~\ref{opeapprox} represents the
energy of the intermediate state. It is proportional to the width of the rectangle and is given by
\be \la{energ}
E = 2 \Gamma_{\rm cusp} ( 2 \sigma_0 + \sigma_s) ~,~~~~~~~\sigma_s \equiv \hat \sigma_1 + \hat \sigma_2\,.
\ee
Thus, we have essentially the same setup that we were using when we talked about the Wilson loop. In
fact, it is the same correlator, except that here we are thinking in terms of an ordinary OPE and we
are only concentrating on the terms going like $e^{ - \tau E}$ where $E$ is the twist of the operator.
In \nref{energ} we ignored the possibility of adding a $\hat \sigma_i$ independent constant. This
is included in the function $\tilde g$. In conclusion, \nref{thirdlineF} is what we expect for the
OPE as $z \to 0$ with $\bar z \sim 1$ from our point of view.

 The OPE  was analyzed in great detail by Dolan and Osborn in
 \cite{Dolan:2004iy}.
 { They considered the asymptotic behavior of the four-point correlator for}
 $z \to 0$ with $\bar z$ fixed,
 and isolated the contributions from
 the conformal tower with the conformal spin $S$.
 Each conformal tower has an anomalous dimension which
 uniquely specifies it and is given { for large spin} by
 $E = \Delta - S = 2 \Gamma_{\rm cusp} \log S  $ \cite{Korchemsky:1992xv},
  where we ignored the subleading constant that is
 independent of the spin.
 For large $S$  the OPE takes the form
 \beqa  &&  \label{opeosborn} 
 G_4\sim 
 \sum_S  z^{ \Delta - S \over 2 } H_S(\bar z) = \sum_S z^{   \Gamma_{\rm cusp} \log S } H_S(\bar z )\,,
 \eeqa
{ with $H_S$ given in terms of a hypergeometric function,}
 \beqa
  \la{osbr}
&&  H_S(\bar z)  = {  \Gamma(S +2 ) \Gamma(S + 3) \over \Gamma(2 S + 4) }
 {\bar z }^S  F(a,a,2a; \bar z) ~,~~~~~a = S +3\,.
 \eeqa
 We now compare \nref{opeosborn} with \nref{thirdlineF}. We need to translate between $\bar z$ and the spin $S$
 and the variables $\sigma_s$ and $\sigma_0$.
 From \nref{CRu} we find that $ (1 - \bar z ) \propto e^{ - 4 \sigma_0}$. Comparing
    the term involving the  anomalous dimensions (the $z$ dependent term)
    in \nref{opeosborn} with \nref{thirdlineF} we find
 \be
 \log S = 2 \sigma_0 + \sigma_s\,.
 \ee
 In the large $S$ limit we can convert the sum over $S$ to an integral and then
 turning this into an integral over $\sigma_s$ we obtain the following measure factor
 $\sum_S \rightarrow \int dS= e^{2\sigma_0} \int e^{\sigma_s}
 d\sigma_s$.
 After multiplying \nref{osbr} by the tree level $(1-\bar z)$ to remove the tree level singularity we
 find
 \be  \label{osbrfin}
 \frac{G_4}{G_4^{\rm tree}}\sim
  \lim_{\sigma_0 \to \infty } (1 - \bar z )  \sum_S  z^{ { \Delta - S  \over 2 } } H_S(\bar z)  \propto
  \int d\sigma_s \, z^{ \Gamma ( 2 \sigma_0 + \sigma_s) }  e^{2 \sigma_s} K_0( 2 e^{\sigma_s })\,,
  \ee
 where we used
 %
 %
 \footnote{This limit can be understood as follows. $H_S(\bar z)$ satisfies a simple quadratic differential equation on $\bar z$, where
 $S$ plays the role of a parameter. Making a change of variables $\bar{z}=1-\epsilon,S = \frac{x}{\sqrt{\epsilon}}$ we can see this equation
 as an equation for $H(\epsilon,x)$. Assuming a small $\epsilon$ behavior of the form $H(\epsilon,x)= \frac{h(x)}{\sqrt{\epsilon}}+...$, which is easy to show,
 it can be seen that $h(x)$ satisfies a simple differential equation whose solution is $x K_0(2x)$. }
 \be \la{limos}
 (1 - \bar z) H_S(\bar z) \propto e^{-2 \sigma_0} e^{\sigma_s} K_0( 2 e^{\sigma_s }
 ),~~~~~~~ {\rm for} ~~~~~ \sigma_0 \gg 1\,.
 \ee
 Comparing the $z$ independent part of the integrand in \nref{osbrfin} with \nref{thirdlineF} we
 find that
 \be
 \int d\hat \sigma_1 u(\hat \sigma_1) u(\sigma_s - \hat \sigma_1 ) = e^{2 \sigma_s} K_0( 2 e^{\sigma_s })\,.
 \ee
 Fourier transforming on both sides gives
 \be
  [v (k) ]^2 \propto \int d \sigma_s e^{ i k \sigma_s } e^{2 \sigma_s} K_0( 2 e^{\sigma_s } )
  \propto \left[\Gamma\left( 1 + { i k \over 2 } \right)\right]^2\,.
  \ee

Returning to our problem, we can now evaluate the factor
$J$ in \nref{Jfour}
 \beqa\notag
  J & \sim &  \left[\Gamma\left( 1 + {   \Gamma_{\rm cusp} \log z  \over 2 } \right)\Gamma\left( 1 + {   \Gamma_{\rm cusp} \log (1-z)  \over 2 } \right)\right]^2
  \\ \label{finj}
  &
=&    e^{ -{ \gamma_{\rm E} } \Gamma_{\rm cusp} (\log z+\log(1-z)) }
 \left\{  1 + { \pi^2  (2\Gamma_{\rm cusp})^2 \over 96} \left[ ( \log z )^2+( \log (1-z) )^2\right]   + \cdots \right\}
\eeqa 
This
computation is only trustworthy to second order in $\Gamma_{\rm
cusp}$. The resulting expression for $J$ reproduces precisely double
logarithms in the second line of \nref{secondline}. The term involving Euler's
constant $\gamma_{\rm E}$ looks like a contribution to the function $\tilde
g$. At one loop we know that there is no single log. Thus this
term must be canceled by a constant contribution to the energy
of the particle, which could come from a simple shift in our
origin of $\sigma$ when we considered the energy eigenstates.
In fact, we can include these constant contributions to $\Delta
-S =  2 \Gamma_{\rm cusp}(\log S+\gamma_{\rm E})- B$. There is
a finite piece involving $\gamma_{\rm E}$ which precisely
cancels the $\gamma_{\rm E}$ in
 \nref{finj}. There is, however a two loop contribution $B$ which precisely matches the
   two loop logarithm in the first line of \nref{secondline}.

Note that the above two-loop computation of $J$  can now be used to write the factor $J$ for any $n$ point correlation function { involving the same BPS operators as in the four-point correlator \p{defco}}
\be
J = \prod_{i=1}^n \check v\left( { \Gamma_{\rm cusp} \over 2 } \log u_{i,i+1} \right) ~,~~~~~~~~~ \check v(w)  \equiv
\Gamma( 1 + w  )e^{ {\gamma_{\rm E} w   } }\,.
\ee
Similarly, the function $\tilde g$ in \nref{finexpan} is equal, up to two loops,  to
\be
\tilde g = - { B \over 2 }\,.
\ee

\section{Conclusions}

In this paper we have considered a multiple light-like limit of correlation function which leads to
Wilson loops. Namely, we started with a correlation function  of local operators $\langle {\cal O}(x_1) \cdots
{\cal O}(x_n) \rangle$ with all points space-like separated. Then we considered the limit where
$x_{i,i+1}^2 \to 0$ but with $x_{i}^\mu \neq x^\mu_{i+1}$, so that the points are separated along a
null direction. In this limit the operators are sitting at the vertices, or cusps, of a polygon with $n$
null edges. As we approach the limit, the correlator develops a singularity that is due to the exchange
of a fast moving particle from one operator to the next.   These fast moving particles define the null polygonal frame which is the source of a color electric field.
If all particles in the theory are in the adjoint representation, this color electric field is in the adjoint.
  This procedure allows us to find the Wilson loop with null polygonal edges as a limit of
the correlator.

In order to extract the Wilson loop, we needed to isolate the contribution from the fast propagating
particles. The easiest way to do this is to dimensionally regularize the correlator and then take the limit
where the distances $x_{i,i+1}^2 \to 0$, with the regulator scale held fixed. In this limit the
particles are propagating in a free theory and their contribution is the same as that of the tree-level
correlator in the same limit. Each particle contributes with a simple factor of $1/x_{i,i+1}^2 $.
Once this factor is extracted { from the correlator} we get the Wilson loop in dimensional regularization \nref{intheadj}.
 Of course, one disadvantage
of this approach is that we need to compute the correlators in dimensional regularization.
 But this is definitely something that one can do.
  For the particular case of ${\cal N}=4$ SYM we demonstrated how to do it at one loop.

One can also take the light-like limit purely in the four dimensional theory. In that case the particles
that propagate are ``dressed'' by the interactions and the approach to the limit is a bit more subtle.
We have untangled these subtleties and arrived at the relation  \nref{OPEWilson} that explains clearly how the limit is approached.
The cusp divergences of the Wilson loop appear as $\propto \Gamma_{\rm cusp} \log x^2_{i,i+1} \log x^2_{i-1,i}$
terms in the exponential. This is a simple and expected factor.
The more  subtle terms involve contributions due to the fast propagating
particles. The factor of the form $\prod 1/(x^2_{i,i+1} )^{ 1 + \tilde g(\lambda)}  $
 arises due to corrections to the ``energies'' of
 the propagating particles.  There is however a more subtle factor that comes from the 
{ back reaction} of
 the color electric field on the propagation of the particle. This is the $J$ factor in \nref{Jgendef}.  { We have computed $J$ explicitly at two loops in  \nref{finj} for the four-point correlator.}
 The extraction of the  Wilson loop from the correlation function can be done in a way that preserves
  conformal symmetry throughout. This is done by extracting the various singular terms via  cross
  ratios. What remains is a renormalized Wilson loop which is explicitly conformally invariant
  \nref{finexpan}.
  The precise expression depends on how we extracted the divergent terms. If we extract them in terms
  of a special kind of cross-ratios, then the Wilson loop is defined by subtracting a very specific
  function, called ``BDS-like'', from the finite piece of the Wilson loop computed in the dimensional regularization.
  The finite piece of the dimensionally regularized Wilson loop is not conformally invariant but it obeys
the anomalous Ward identity   \cite{Drummond:2007aua}. The BDS-like expression \cite{AGM} is the unique  way to solve this anomaly by considering
  a function of only next-to-nearest neighbor distances, $x_{i,i+2}^2$. This procedure works and is well defined
  when the number of sides is not a multiple of four, $n \not = 4 k$. We have not treated the $n= 4 k$
   case here
  and we leave that for the future. However, we did check that the correlator reproduces
  the BDS $-$ (BDS-like)  expression at one loop.

This basic connection between correlators and polygonal Wilson loops should be true for general conformal
gauge theories in any space-time dimensions.

This expansion could  also be done for ${\cal N}=4 $ super-Yang-Mills at strong coupling by using
strings in $AdS_5\times S^5$. We expect it to work in a similar fashion. In fact, a
closely related computation was done in appendix B.1 of \cite{AGMSV}, where the large spin limit of single trace operators
was considered.

In planar { gauge} theories the adjoint Wilson loop can be viewed as a product of a fundamental and an anti-fundamental Wilson loops.  In theories with dynamical fundamental fields one can of course get directly the
Wilson loop in the fundamental by considering mesonic operators. Of course, one should also face the
issue that the Wilson loop can be screened. If one is doing perturbation theory, or considering only the
planar theory,  then one does not
need to worry about screening issues.

In this paper we assumed that the operators { entering the correlator} were such that the tree-level contribution { to the correlator} would always lead
to a particle propagating from $x_i$ to $x_{i+1}$. One could easily imagine cases where this would not
be the case. For example, one could have a charged chiral primary operator,
such as ${\rm Tr}[ ( \phi_1 + i \phi_2)^2]$
at $x_i$ and $x_{i+1}$. It is possible that one could slightly modify the discussion here so that
these cases are also covered.   In particular, one would have to understand whether we have the
exchange of any particle between these two operators at higher loop level, etc.

In a parallel publication \cite{EKS} an alternative way
to take the light-cone limit is proposed. The correlators are
 first computed in four dimensions, and are then regularized by
  a ``dual infrared" dimensional regulator. The limit produces the
   MHV gluon scattering amplitudes in the dual momentum space.

\section*{Acknowledgments}

We would like to thank Nima Arkani-Hamed and Vladimir Braun  for interesting discussions.  GK is grateful to Vladimir Bazhanov  and to the Australian National University, Canberra,
and ES is grateful to Nima Arkani-Hamed and the Institute for Advanced Study, for warm hospitality at various stages of this work.  This work was supported
in part by the French Agence Nationale de la Recherche under grant
ANR-06-BLAN-0142. This work was supported in part by   U.S.~Department of Energy
grant \#DE-FG02-90ER40542.

\appendix

\section*{Appendices}

\section{Coordinate changes}
 \label{coordchanges}

We have coordinates $Z^M = (Z^{-1},Z^0,Z^1,Z^2,Z^3,Z^4)$, obeying $0=Z^2 = - (Z^{-1})^2 - (Z^{0})^2+(Z^{1})^2 +
\cdots +(Z^4)^2 $. It is also convenient to introduce the following rotated coordinates
\beqa
  \hat  Z^{-1} &=& { Z^0 + Z^{-1}  \over \sqrt{2} } ~,~~~~~ \hat  Z^0 = {  Z^0 - Z^{-1} \over \sqrt 2}
 ~,~~~~~~~ \hat  Z^{i }=Z^{i }  ~,~~~~~~~(i=1,2,3,4)\,.
\eeqa
We then define the ordinary coordinates of $R^{1,3}$ as
\be
x^\mu = { \hat Z^\mu \over G } ~,~~~~~~~~~~~~  G = \hat  Z^{-1} +  \hat  Z^{4}  ~,~~~~~~~ (\mu =0, 1,2,3)\,.
\ee
We also introduce the $AdS_3\times S^1$ coordinates via
\begin{align}\notag
& \tilde  G  \equiv  \sqrt{ (Z^3)^2 + ( Z^4 )^2 } \,, && e^{ i \varphi } = { Z^3 + i Z^4 \over  \tilde G }\,,
\\ \label{ztau}
& { Z^0 \pm Z^1 \over   \tilde G} =  \cos \alpha\, e^{ \pm \gamma} \,, &&
{ Z^{-1} \pm Z^2 \over   \tilde G} = \sin \alpha\, e^{ \pm \chi}  \,.
\end{align}
We can think of the functions $G$ or $\tilde G$
 as specifying a slice through the projective lightcone $Z^2=0$. The metric
can then be written as $ds^2 = dZ\cdot dZ =  G^2 ds^2 $ where $ds^2$ is the metric in $AdS_3\times S^1$ or
the metric of $R^{1,3}$ depending on whether  we pick  $G$ or $\tilde G$ in this formula.

These relations map the square polygon whose vertices are at
$x^\mu = (1, \pm  {1 \over \sqrt{2}}, 0,0), ~ (-1,0,
\pm { 1 \over \sqrt{2} } , 0) $ to a polygon whose null lines are at infinity in $\gamma$ and $\chi$.
The propagator in any coordinate system is given by
\be
\langle \phi (1) \phi(2) \rangle = - { G(1) G(2) \over 2  Z_1\cdot Z_2} ~,~~~~~~~
 \langle \phi (1) \phi(2) \rangle_{\rm flat} =
{ 1 \over x_{12}^2 } = - {  G(1)  G(2) \over 2 Z_1\cdot Z_2}\,.
\ee
This formula can also be used to express distances in $R^{1,3}$ in terms of the $AdS_3 \times S^1 $ coordinates
\beqa \notag
x_{12}^2 &=&
 - { 2 Z_1 \cdot  Z_2 \over  G(1)  G(2) } = - 2 {  Z_1\cdot  Z_2  \over \tilde G(1) \tilde G(2) } \lr{
 \tilde G(1) \tilde G(2) \over G(1) G(2)  } =
\\
&=&2 \big[ \sin \alpha_1 \sin \alpha_2 \cosh \chi_{12} +   \cos \alpha_1 \cos
\alpha_2 \cosh \gamma_{12} - 1 - \cos \varphi_{12}\big] \Omega(1) \Omega(2)
\eeqa
with
\beqa \label{Omegafa}
\Omega &=& { \tilde G   \over G  } =  { 1 \over { 1 \over \sqrt{2} } (  \sin \alpha \cosh \chi + \cos \alpha \cosh \gamma ) +
\sin \varphi }
\eeqa
We can also introduce $\gamma = \tau + \sigma$, $\chi = \tau - \sigma$  and $\alpha = \pi/4 + \beta$.
Now the distance between two points at the same values of $\sigma$, $\sigma_1 = \sigma_2 =\sigma_0 \gg 0$, at
$\alpha_1 =\alpha_2 = \pi/4$  and two different but large
values of $\tau$ is
\be
x_{12 }^2 \sim   2  { e^{ |\tau_{12} |}  \over   e^{ |\tau_1| + \sigma_0 } e^{|\tau_2| + \sigma_0 } }\,.
\ee
If $\tau_{1,2} = \pm \tau_0$ then we see that $x_{12}^2 \propto e^{ - 2 \sigma_0 }$.

If we have four arbitrary and space-like separated points $x_i$, ($i=1,2,3,4$) we can map them via
a conformal transformation to a rectangle in the $\tau$ and $\sigma$ plane at $\beta= \varphi=0$.
The rectangle is characterize by its width and height, $\Delta \tau$ and $\delta \sigma$.
We can then compute cross ratios by writing the coordinates $Z$ in terms of $\tau, \sigma$
as in \nref{ztau} and then computing $\chi_{1234} = { Z_1\cdot Z_2 Z_3\cdot Z_4 \over Z_1\cdot Z_3 Z_2\cdot Z_4 } $, etc.
Using this we get to \nref{CRu}.

\section{Calculating the integral $I_\ep(x)$}  \label{apA}

Here
we evaluate the function $I_\epsilon$ which appears in \p{vertdi}.
To simplify the notation, we use translation invariance to set the
external point $x_0=0$ and relabel the other points as follows:
\begin{equation}\label{ap1}
T^{\mu\nu}(x_1,x_2)=    \partial_1^{[\mu}\partial_{2}^{\nu]}\int \frac{d^{4-2\ep} x_{3}}
{(-x_{13}^2 x_{23}^2 x_{3}^2)^{1-\epsilon}}\,.
\end{equation}
Introducing Feynman parameters, we find
\begin{eqnarray}  \label{ap2}
T^{\mu\nu}(x_1,x_2) &=& \partial_1^{[\mu}\partial_{2}^{\nu]} \prod_{i=1}^3 \int_0^\infty \frac{(-i)^{1-\ep}ds_i s^{-\ep}_i}{\Gamma(1-\ep)} \int d^{4-2\ep} x_{3} \e^{-is_1 x^2_{13} - is_2 x^2_{23} - is_3 x^2_3} \\
  &=& \frac{4 i^{2\ep}\pi^{2-\ep} x^{[\mu}_1 x^{\nu]}_2}{\Gamma^3(1-\ep)} \int_0^\infty \frac{ ds_1ds_2ds_3 (s_1 s_2 s_3)^{1-\epsilon}}{  (s_1+s_2+s_3)^{3-\epsilon}}\exp\lr{-i\frac{x_{12}^2 s_1 s_2+x_{1}^2 s_1 s_3+x_{2}^2 s_2 s_3 }{s_1+s_2+s_3}}\,. \nn
\end{eqnarray}
Comparing to \p{vertdi} and changing the integration variables
$s_i=\lambda \beta_i$ (with $\lambda>0$ and $\sum_i\beta_i=1$), and after the elementary integration over
$\lambda$ we obtain the following integral representation of the function $I_\epsilon$:
\begin{align}
I_\epsilon (x_1,x_2) = -4i \pi^{2-\ep} \frac{\Gamma(3-2\ep) }{\Gamma^3(1-\ep)} \int_0^1 \frac{ [d\beta]_3\, (\beta_1 \beta_2 \beta_3)^{1-\epsilon}}{(-x_{12}^2 \beta_1\beta_2+x_{1}^2 \beta_1\beta_3+x_{2}^2 \beta_2 \beta_3 )^{3-2\epsilon}}\,,
\end{align}
with $[d\beta_3]=d\beta_1d\beta_2d\beta_3\delta(1-\beta_1-\beta_2-\beta_3)$.
Changing once again the variables according to
$$
\beta_1= \frac{s}{1+z}\,,\quad \beta_2=\frac{1- s}{1+z}\,,\quad \beta_3=\frac{z}{1+z}
$$
with  $0\le s \le 1$ and $0\le z <\infty$,  we obtain
\begin{align}
I_\epsilon (x_1,x_2)
&=  -4i \pi^{2-\ep} \frac{\Gamma(3-2\ep) }{\Gamma^3(1-\ep)}\int_0^\infty \frac{dz z^{1-\ep}}{ (1+z)^{3-\ep}} \int_0^1\frac{ds\, (s\bar s)^{1-\ep}}{ [- x_{12}^2 s \bar s - z (x_1^2 s + x_2^2 \bar s)]^{3-2\ep}}
 \label{A4}
\end{align}
with $\bar s=1-s$.

We are interested in the leading asymptotic behavior of \re{A4} for $x_{12}^2\to 0$. In this limit, replacing  $z\to z x_{12}^2 s\bar s/(x_1^2 s + x_2^2 \bar s)$
and noticing that $x_{1}^2 s + x_{2}^2 \bar s =  (x_{1} s + x_{2} \bar s )^2$ for $
x_{12}^2=0$, we find
\begin{align}\label{intI'}
I_\epsilon (x_1,x_2)& \to -4i \pi^{2-\ep}\frac{\Gamma(2-\ep)}{\Gamma^2(1-\ep)}\, (-x_{12}^2)^{-1+\epsilon}\,
 \int_0^1 ds\, \left[-(x_{1}  s + x_{2}  \bar s)\right]^{-2+\epsilon} \qquad \mbox{for $x_{12}^2\to 0$}
\end{align}
as announced in \p{intI}.
Integrating with respect to $s$ yields
\begin{align}\label{II}
I_\epsilon (x_1,x_2)\sim - \frac{4i \pi^{2-\ep}}{\Gamma(1-\ep)}\, (-x_{12}^2)^{-1+\epsilon}\, \frac{(-x_{1}^2)^{-1+\epsilon}-(-x_{2}^2)^{-1+\epsilon}}{x_{1}^2-x_{2}^2}\,.
\end{align}
For $\epsilon\to 0$ we recover \p{I0}.

\section{Scalar propagator on the light cone}\label{App-scalar}

In this appendix, we work out the first few terms in the light-cone expansion
of the scalar propagator in an external gauge field, Eqs.~\re{ansatz} and \re{OPE}. We recall the basic assumption that the light-cone limit is taken so that $x^2\mu^2 \ll 1$. This allows us to neglect the presence of the dimensional regularization cutoff $\mu^2$ in all formulas.

We start with the definition \re{D-eq} and employ the proper time technique to write
the solution to \re{D-eq} as (for $y=0$)
\begin{align}\label{pro'}
S(x,0;A) =  \vev{x|\frac1{iD^2}|0}=\int_0^\infty ds\, \vev{x|\e^{-is D^2}|0}\,.
\end{align}
Here $D^2=D_\mu^2$ and $[D_\mu]^{ab} = \partial_\mu \delta^{ab} + g f^{abc} A_\mu^c$ is the gauge covariant derivative in the adjoint representation. The expansion
of $\vev{x|\e^{-is D^2}|0}$ in powers of the gauge field strength $F_{\mu\nu}$ looks as \cite{Balitsky:1987bk}
\begin{align}\label{non-loc}
\vev{x|\e^{-is D^2}|0} &= \vev{x|\e^{-is \partial^2}|0}\bigg\{ [x,0]+s\int_0^1 du \, u\bar u\,
[x,ux] x_\nu D_\mu F^{\mu\nu}(ux) [ux,0]
\\ \notag
& + 2is \int_0^1du\,\bar u \int_0^1dv\, v\, x_\lambda x_\rho  [x,ux]F^{\lambda\xi}(ux) [ux,vx] F^{\rho}{}_\xi(vx)[vx,0]+ O(s^2)\bigg\}\,,
\end{align}
where $[x,y]\equiv P\exp\lr{i\int_{y}^x dz\cdot  A(z)}$ stands for a Wilson line
in the adjoint representation connecting points $x$ and $y$, and $\bar u=1-u$.

The first factor on the right-hand side of \re{non-loc} can be easily evaluated by going
to the momentum representation
\begin{align}
\vev{x|\e^{-is \partial^2}|0} = \int \frac{d^4 p}{(2\pi)^4}\, \e^{-ipx} \e^{is p^2} \sim s^{-2} \e^{-i x^2/(4s)}\,.
\end{align}
Its substitution into \re{pro'} yields the free propagator
\begin{align}
S^{\rm tree}(x) =   \int_0^\infty ds\, \vev{x|\e^{-is \partial^2}|0} = - \frac1{(2\pi)^2 x^2}\,.
\end{align}
Likewise, substituting \re{non-loc} into \re{pro'} and performing the
$s-$integration term by term, we find that the expansion of \re{non-loc} in $s$ can be translated into a similar expansion of $S(x,0;A)$ in $x^2$:
\begin{align}\label{S-app}
S(x,0;A) = - \frac{ 1}{(2\pi)^2 x^2}P\exp\lr{ig\int_{0}^x dz\cdot A(z)}
 \left[ 1+ x^2 \lr{\mathcal{O}_1(x) + \mathcal{O}_2(x)}+O(x^4) \right]\,,
\end{align}
where $\mathcal{O}_1$ and $\mathcal{O}_2$ are the so-called light-ray operators
 \cite{Balitsky:1987bk,Mueller:1998fv}
\begin{align}
\mathcal{O}_1 &\sim \int_0^1 du \, u\bar u\,
[0,ux] x_\nu D_\mu F^{\mu\nu}(ux) [ux,0]
\\ \notag
\mathcal{O}_2 &\sim 2i  \int_0^1du\,\bar u \int_0^1dv\, v\, x_\lambda x_\rho  [0,ux]F^{\lambda\xi}(ux) [ux,vx] F^{\rho}{}_\xi(vx)[vx,0]\,.
\end{align}
These operators can be thought of as generating functions for twist-two operators, e.g.
\begin{align}
\mathcal{O}_1\sim \int_0^1 du \, u\bar u\,\e^{u (x\cdot D)}
 x_\nu D_\mu F^{\mu\nu}(0)
 = \sum_{N\ge 0}\frac{N+1}{(N+3)!} (x\cdot D)^N x_\nu D_\mu F^{\mu\nu}(0) \,.
\end{align}
We conclude from \re{S-app} that the leading contribution to the scalar propagator for $x^2\to 0$ comes from the first term in the square brackets corresponding to the identity operator with twist zero.

\section{A Factor of two}
\label{FactorTwo}

In this appendix we explain the relative factor of two which arises in front of double
logarithmic terms when we compute the Wilson loop in dimensional regularization versus the
case when we compute the limit of the correlation functions and set the distances equal to
$x_{i,i+1}^2 = \mu^{-2}$.
For the most divergent terms we get
\beqa
\left. \log \langle O(1) \cdots O(n) \rangle \right|_{x_{i,i+1}^2 = \mu^{-2} }  & \sim &  - n
{\Gamma_{\rm cusp} \over 2 } ( \log \mu^2 )^2\,,  \label{corrcusp}
\\
\log W^{\rm adj}_{\rm dim-reg} & \sim &  -n {\Gamma_{\rm cusp} \over 4 } ( \log \mu^2 )^2\,,
 \label{dimregcusp}
\eeqa
where $n$ is the number of cusps.
Naively one would have expected that this most divergent term should have matched.

The relative factor of two has a simple geometric
explanation if we go to the coordinates we introduced in
appendix \ref{coordchanges}. For simplicity consider the case of the square, which contains four cusps.

\begin{figure}[t]
\begin{center}
\includegraphics[width=80mm]{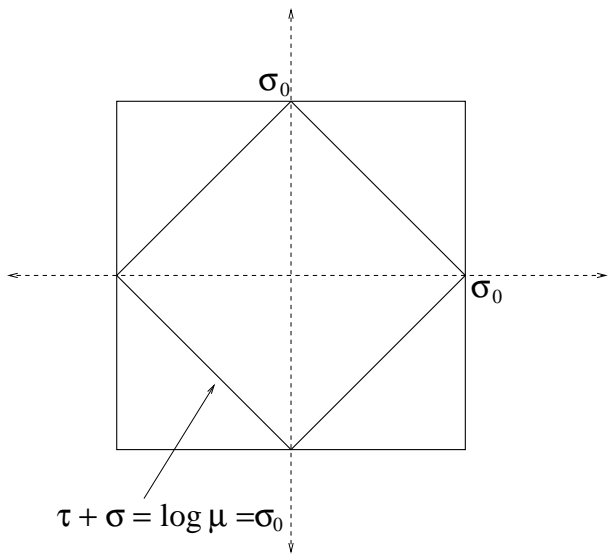}
\end{center}
\caption{ Origin of a factor of two due to the different ways of regularizing the contribution
of the color electric flux. In the case of correlation function we have operators sitting at the
vertices of the outer square. Fast propagating particles define the outersquare and the contribution
of the flux is the area of this square. In dimensional regularization we end up with a coupling
constant that depends on scale in these new coordinates. The coupling switches itself off outside
the small square. Thus the contribution is just the area of the small square.
} \la{factors}
\end{figure}

Let us set the operators in the correlation function at the
vertices of a square at $\tau = \pm \sigma = \pm \sigma_0$.
 Then one finds that the distances are
\begin{equation} \label{coordmu}
x_{12}^2 \approx e^{-2\sigma_0} \sim \mu^{-2},~~~~x_{23}^2 \approx e^{-2 \tau_0} \sim \mu^{-2}  ~,~~~~
\tau_0 = \sigma_0
\end{equation}
The total area of the rectangle is $4 \sigma_0^2  $. This agrees with \nref{corrcusp}, for $n=4$, once
we use \nref{coordmu}.

We now consider a Wilson loop in
the dimensionally regularized theory whose cusps are light-like separated.
Under the conformal mapping the edges of the null square polygon are sent to $\tau = \pm \infty $ and $\sigma = \pm \infty$.
The coupling constant becomes $\tau$ and $\sigma$ dependent because the theory is not
conformal invariant. One can do conformal transformations in a non-conformal theory if
one remembers to change the coupling constants that violate the conformal symmetry.
In order to find the new coupling constant we can do the following. The coupling in the
dimensionally regularized theory has a scale $\mu$. For our purposes,
we can replace $\mu \to \mu\,\Omega$, where
$\Omega$ is given in \nref{Omegafa} and is equal to $\Omega \sim e^{ - |\tau| - |\sigma| }$
for large $\tau$ and $\sigma$. Thus we end up with a coupling that depends on the position
and it is such that it goes to zero for large $\tau, \sigma$. Thus, as expected,
 we get a finite answer for the
contribution of the electric field, since the energy in the electric flux is proportional to the
coupling. The coupling is turning itself off when $\mu \Omega \sim 1$. Thus we
find that $|\tau| + |\sigma| \leq  \log \mu = \sigma_0$. This gives the diagonal
lines in Figure~\ref{factors}. These  produces a square of size
$ 2 \sigma_0^2$,  in agreement with  \nref{dimregcusp}, with $n=4$.

This explanation is completely general and is valid whenever the formulas  \nref{corrcusp} and \nref{dimregcusp} are valid. This same factor of two was observed in the closely related problem of QCD form factors
in \cite{Korchemsky:1988hd}.
%


\end{document}